\documentstyle[prl,floats,aps,epsf,psfig,eqsecnum]{revtex}
\newcommand{\statewidth}{3in}
\newcommand{\figwidth}{6in}
\begin{document} 
\title{Magnon Wave-function and Impurity Effects in S=1 Antiferromagnetic
Chains: A Large-n Approach} 
\author{Alistair Savage$^1$\cite{AS} and Ian Affleck$^{1,2}$}
\address{$^1$Department of Physics and Astronomy and $^2$Canadian
Institute for Advanced Research, \\ University of British Columbia,
Vancouver, BC, Canada, V6T 1Z1}
\maketitle 
\begin{abstract}  
A large-$n$ approximation to the $S=1$ antiferromagnetic
chain, using the symmetric tensor representation and its conjugate, is
developed to order $1/n$ in order to calculate the magnon wave-function
and to study the effect of modifying the exchange coupling from $J$ to
$J'$ on a single link.  It is shown that a magnon boundstate exists below the
Haldane gap for arbitrarily small negative $J'-J$ but only above a certain
critical value of $J'-J$ for positive values.  In the former case the binding
energy vanishes as $(J-J')^2$.
\end{abstract}
\section{Introduction and Review}
Many features of $S=1$  Heisenberg antiferromagnetic chains have been
understood in considerable detail using a combination of numerical
techniques, approximate field theory methods and a model with bilinear
plus biquadratic exchange for which the exact groundstate is a simple
valence bond solid (VBS).\cite{Affleck1}  There is considerable interest in
the first excited state, the triplet magnon, separated from the
groundstate by the Haldane gap.  However, this state is not known exactly even
for  the VBS model.

Investigations have been made of the Heisenberg model
with the  exchange coupling, $J$, modified to a different value, $J'$, on a
single link.  It can be seen that a magnon boundstate exists, localized
near the modified link, whose energy goes to zero in the limit $J'/J<<1$
or $J'/J>>1$. 
However controversy remains over whether this boundstate continues to
exist for $|J'-J|<<J$.  The numerical results in \onlinecite{Sorensen}
suggest that a critical value of $J'-J$ is required, for either sign of $J'-J$,
in  order for the boundstate to exist.  On the other hand, the more
detailed numerical analysis 
 of \onlinecite{Mallwitz} suggest that this is only true for
$J'>J$  while for $J'<J$ a magnon boundstate exists for arbitrarily small
values of $J-J'$.
There is a well known  theorem of one dimensional quantum mechanics  which
states that an arbitrarily weak attractive potential produces a
boundstate. (For a general and rigorous version of this theorem see
\onlinecite{Simon}.)  Since the magnon behaves in many ways like an ordinary
stable massive particle, one might expect that this theorem should be
applicable to the modified link problem, but this connection has not
been  established.  

The valence bond solid state was in fact encountered
earlier\cite{Affleck2} as the groundstate of the bilinear exchange
model in a certain large-$n$ limit. This is defined by placing the
symmetric 2-tensor representation of $SU(n)$ on the even sites and its
conjugate representation on the odd sites.  This is sometimes referred to
as a type of ``fermionic'' large-$n$ limit because it arises in a model with
2 ``colors'' of fermions and $n$ ``flavors'' with a projection onto color
singlets on each site with 2 particles on the even sites and 2 holes
($n-2$ particles) on the odd sites.  This must be distinguished from
various other large-$n$ limits including the fermionic case with $n/2$
fermion of each color on each site and the bosonic model with $n$ bosons
on even sites and $n$ anti-bosons on odd sites.  In
\onlinecite{Affleck2} the model was solved only to leading order in
$1/n$.  In this order, there remains a 3-fold groundstate degeneracy which is lifted in higher
orders. The VBS groundstate is degenerate, to leading order in $1/n$, with the
two fully dimerized groundstates \cite{Affleck2}. The Haldane gap, of $O(1)$
was calculated in \onlinecite{Affleck2} but a determination of the
corresponding magnon  wave-function requires going to order $1/n$, due to
an infinite degeneracy at lowest order.  

The purpose of this paper is to develop this large-$n$ approximation to
next order in $1/n$.  We show that indeed the VBS state is the actual
large-$n$ groundstate, the dimerized states having energy higher by
$O(1/n)$. [It was erroneously claimed in \onlinecite{Affleck2} that this
splitting only occurs at $O(1/n^2)$.]  We also solve for the magnon
wave-function in the large-$n$ approximation.  In this approximation it
corresponds to a type of soliton-antisoliton boundstate held together by a
linear potential.  Our
results allow us to estimate the accuracy of the large-$n$ approximation for
the physical case $n=2$.  The results are somewhat encouraging.
In any event, this approximation is useful because it qualitatively
captures the
essential physics.   As an application, we show that the large-$n$
approximation allows for a simple analytic treatment of the
single modified link problem. We show  that a boundstate
exists for arbitrarily small negative $J'-J$ but for positive $J'-J$
only when $J'-J$ is greater than a critical value, in agreement with
Mallwitz and Wang\cite{Mallwitz} with the binding energy
vanishing as $(J-J')^2$ in the $J'<J$ case.  This result follows
essentially from the theorem of one-dimensional quantum mechanics
that an arbitrarily weak attractive potential always produces a
boundstate. The
simplicity and generality of the argument suggests that it may remain
true for all n.  

For general $n$, the states on even sites are labeled by a pair of
symmetric $SU(n)$ particle indices:
\begin{equation} |^{ij}>=|^{ji}>\end{equation}
and on odd sites by a pair of symmetric hole indices:
\begin{equation} |_{ij}>=|_{ji}>.\end{equation}
For all $n\geq 3$ even and odd sites are inequivalent.  However, in
the n=2 case they are equivalent with the 3 states on each link
corresponding to the usual $S^z$ eigenstates:
\begin{eqnarray}
|^{11}>&=&|_{22}>=|S^z=1>\nonumber \\
|^{12}>&=&|_{12}>=|S^z=0>\nonumber \\
|^{22}>&=&|_{11}>=|S^z=-1>.\end{eqnarray}
For the 2-site model the Hamiltonian acts as:
\begin{equation} H|^{ij},_{kl}>=-(1/n)[\delta^i_k|^{mj},_{ml}>+
\delta^i_l|^{mj},_{km}>+\delta^j_k|^{im},_{ml}>+\delta^j_l|^{im},_{km}>].
\label{Hdef}\end{equation}

For the case $n=2$,
\begin{equation} H = \vec S_1 \cdot \vec S_2 - 1 \label{Hshift} \end{equation}
Throughout this paper we use the Einstein summation convention for
a pair of repeated indices one upper and one lower.  All $SU(n)$
singlet    states for the chain can be written with all upper indices
contracted with lower indices.  States with uncontracted upper and lower
indices can    be projected into irreducible representations of $SU(n)$ by the
usual    processes  of symmetrization and subtracting of traces.  

For the 2-site model it is
quite easy to see that the groundstate is the $SU(n)$ singlet
$|^{ij},_{ij}>$ with energy -2 in the large-$n$ limit.
The factor of $1/n$ in (\ref{Hdef}) is
canceled by $\delta^i_i=n$.   Now consider a chain of arbitrary
length.   We may represent an arbitrary $SU(n)$ singlet state by drawing
a line    (``valence bond'') between pairs of sites (one even and one odd)
representing each contracted pair of indices.  In the large-$n$ limit
the energy of such a state is simply equal to (-1) times the number of
nearest neighbor valence bonds.  Hence there are three degenerate
groundstates in the large-$n$ limit, each of which has one nearest neighbor
valence    bond per link, on average.  Two of these states are the
dimerized states with a pair of valence bonds between each even site and
the odd site to its  right (or left).  [See Figure \ref{fig:grounddef}.]
The third
degenerate    groundstate is the valence bond solid state with a single
valence bond between    each neighboring pair of sites. [See Figure
\ref{fig:grounddef}].  Only by going
to higher order in $1/n$ will the degeneracy between the dimerized and
VBS groundstates be lifted.  We show in the next section that the VBS
state is    the true unique groundstate once $O(1/n)$ corrections are
included.

In the large-$n$ limit, the lowest excited state has one fewer
nearest    neighbor bond and hence an excitation energy of 1.
Assuming the VBS    wave-function at positive and negative spatial infinity,
we see that    all such states correspond to configurations with a string
of dimers    between
the VBS states and dangling bonds separating the VBS regions from
the dimer regions.  [See Figure \ref{fig:solitondef}.] These dangling bonds 
correspond to solitons between the nearly degenerate VBS and dimer
approximate groundstates.  Such states are labeled by 2 indices, one
upper and one
lower corresponding to the two dangling bonds, $|^i,_j>$.  Subtracting the
trace, $|^i,_j>-(\delta^i_j/n)|^k,_k>$ gives a state in the adjoint
representation of $SU(n)$.  In the $n=2$ case this reduces to the spin-1
(triplet) representation.  To lowest order in $1/n$, all such
soliton-antisoliton ($s\bar s$) states have the same excitation energy,
+1.  It is necessary to go to $O(1/n)$ to split the degeneracy.  Note that,
if the soliton and antisoliton are separated by a distance, $x$, then the
energy will grow as $x/n$ because there is a region of size $x$ which is in
the ``wrong'' (dimerized) groundstate, whose energy is higher by an amount
of $O(1/n)$.  Since the kinetic energy for the soliton and antisoliton is
also $O(1/n)$, it follows that the binding energy is $O(1/n)$.  We essentially
must solve a lattice version of the Schroedinger equation for a particle
in a linear potential.  In general this will give several
soliton-antisoliton boundstates, each corresponding to an adjoint
representation magnon.  The number of stable boundstates grows with
increasing $n$. 

In the next section we study the translationally invariant case,
calculating the groundstate and magnon states to $O(1/n)$.  In Section III
we consider the single modified link.  A preliminary version of Section II
appeared in \onlinecite{Savage}.

\section{Groundstate and Magnon States}
\subsection{Groundstate}
We label the candidate groundstates as follows.

\begin{eqnarray}
|0>&=&|...^{i_1i_2},_{i_2i_3},^{i_3i_4},_{i_4i_5}...>\nonumber \\
|EO>&=&|...^{i_1i_2},_{i_1i_2},^{i_3i_4},_{i_3i_4}...>\nonumber \\
|OE>&=&|...^{i_1i_2},_{i_3i_4},^{i_3i_4},_{i_5i_6}...>\end{eqnarray}

\begin{figure}
\epsfxsize=\statewidth
\centerline{\epsffile{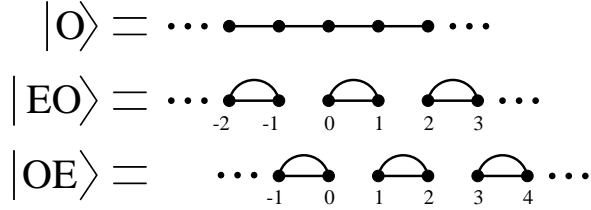}}  
\caption{Candidate Groundstates}
\label{fig:grounddef}
\end{figure}

Thus the $|EO>$ and $|OE>$ states are related by a translation of one site.
$|EO>$ is the state in which the double bonds have even sites on the left
and odd sites on the right and vice versa for $|OE>$
(see Figure \ref{fig:grounddef}).
We expect the energy of the non-symmetric states to be identical and that
of the symmetric state to be different (since no symmetry connects it to the
non-symmetric states).  Note that all three states have one bond per link on
average.  We now show that the VBS state $|0>$ is the true groundstate
for large $n$

\begin{figure}
\epsfxsize=\statewidth
\centerline{\epsffile{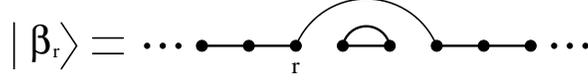}}  
\caption{$\beta$ state}
\label{fig:betadef}\end{figure}

We wish to calculate the energy of the symmetric groundstate including
the first order correction.  This is
\begin{equation} E^{(1)}_0=<0|H|0>/<0|0>\end{equation}
Now, we first consider the action of the Hamiltonian on just two
sites of the symmetric ground state.  That is, we calculate $H_r|0>$
where $H_r$ is defined to be the term in $H$ acting on sites $r$ and $r+1$.  We
consider only the two sites on which $H_r$ acts.
\begin{eqnarray}
H_r|_{i_1i_2},^{i_2i_3}>&=&-(1/n)[\delta^{i_2}_{i_1}|_{mi_2},^{mi_3}>+
\delta^{i_2}_{i_2}|_{i_1m},^{mi_3}>+\delta^{i_3}_{i_1}|_{mi_2},^{i_2m}>+
\delta^{i_3}_{i_2}|_{i_1m},^{i_2m}>]\nonumber \\
&=&-(1/n)[|_{mi_1},^{mi_3}>+n|_{i_1m},^{mi_3}>+
\delta^{i_3}_{i_1}|_{mi_2},^{i_2m}>+|_{i_1m},^{i_3m}>]
\end{eqnarray}
Renaming indices yields
\begin{equation}
H_r|_{i_1i_2},^{i_2i_3}>=-(1+2/n)|_{i_1i_2},^{i_2i_3}>-
(1/n)\delta^{i_3}_{i_1}|_{mi_2},^{i_2m}>\end{equation}
The first term is simply a multiple of the original state.  The second term
consists of a double bond and a delta function.  This delta function serves to
contract the two sites on either side of the two sites considered here thus
producing a $|\beta_{r-1}>$ state (see Figure \ref{fig:betadef}).
Thus the action of the
Hamiltonian on the entire symmetric ground state ($L$ sites) is
\begin{equation}
H|0>=-L(1+2/n)|0>-(1/n)\sum_r|\beta_r>\end{equation}
If we assume that the states $|0>$ and $|\beta_r>$ are orthogonal,
\begin{equation}
<0|H|0>=-L(1+2/n)<0|0>\end{equation}
And so
\begin{equation}
<0|H|0>/<0|0>=-L(1+2/n) \label{Oenergy}\end{equation}
In fact, the states $|0>$ and $|\beta_r>$ are not orthogonal.  However, their
overlap (divided by $<0|0>$) is order $1/n$
(see Appendix \ref{app:overlaps}) and thus the effect of the non-orthogonality
is order $1/n^2$.

We now consider the non-symmetric groundstate.  As above, we wish to calculate
\begin{equation} E^{(1)}_{EO}=<EO|H|EO>/<EO|EO>\end{equation}
which is the energy of the non-symmetric ground state including
the first order correction.  Now, in this case we need to
consider the action of the Hamiltonian
on two inequivalent types of pairs of sites, those possessing a
double bond between each other and
those not.  For a chain of $L$ sites,
there are $L/2$ of each type (assuming L is even).

\begin{figure}
\epsfxsize=\statewidth
\centerline{\epsffile{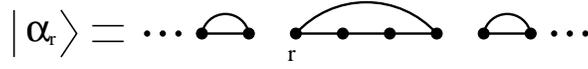}}  
\caption{$\alpha$ state}
\label{fig:alphadef}\end{figure}

First consider $H_r$ with $r$ even.  Then $H_r$ acts on a pair of sites
possessing a double bond between each other.  We need only consider
the two sites on which the Hamiltonian acts.
\begin{eqnarray}
H_r|_{i_1i_2},^{i_1i_2}>&=&-(1/n)[\delta^{i_1}_{i_1}|_{mi_2},^{mi_2}>
+\delta^{i_1}_{i_2}|_{i_1m},^{mi_2}>+\delta^{i_2}_{i_1}|_{mi_2},^{i_1m}>
+\delta^{i_2}_{i_2}|_{i_1m},^{i_1m}>] \nonumber \\
&=&-(1/n)[n|_{mi_2},^{mi_2}>+|_{i_1m},^{mi_1}>+|_{mi_1},^{i_1m}>
+n|_{i_1m},^{i_1m}>]
\end{eqnarray}
We see that all of the states in the above expression are simply
the original state with a change of dummy variables.  Therefore,
\begin{equation}
H_r|_{i_1i_2},^{i_1i_2}>=-2(1+1/n)|_{i_1i_2},^{i_1i_2}>
\end{equation}
Now consider
$H_r$ with $r$ odd.  Then $H_r$ acts on a pair of sites not possessing
a double bond between each other.  Again, we only consider the two
sites on which the Hamiltonian acts.  Remembering that the indices of
the leftmost site are contracted to the left and those of the
rightmost site are contracted to the right,
\begin{equation}
H_r|_{i_1i_2},^{i_3i_4}>=-(1/n)[\delta^{i_3}_{i_1}|_{mi_2},^{mi_4}>
+\delta^{i_3}_{i_2}|_{i_1m},^{mi_4}>+\delta^{i_4}_{i_1}|_{mi_2},^{i_3m}>
+\delta^{i_4}_{i_2}|_{i_1m},^{i_3m}>]
\end{equation}
In each of the states in the above expression, there exists a single
bond between the two sites considered (that is, sites $r$ and $r+1$),
a single bond between sites $r-1$ and $r$ resulting from one of the
original dummy indices on site $r$ remaining and a single bond between
sites $r+1$ and $r+2$ resulting from one of the original dummy indices
on site $r+1$ remaining.  Also, in each case, the delta function
serves to contract the indices on sites $r-1$ and $r+2$.  Thus, the
action of the Hamiltonian on the state $|EO>$ is
\begin{equation}H|EO>=-L(1+1/n)|EO>-(4/n)\sum_r|\alpha_{2r}>\end{equation}
where $|\alpha_r>$ is defined in Figure \ref{fig:alphadef}.
If we assume the states $|EO>$ and $|\alpha_{2r}>$ are orthogonal,
\begin{equation}<EO|H|EO>=-L(1+1/n)<EO|EO>\end{equation}
And so
\begin{equation}<EO|H|EO>/<EO|EO>=-L(1+1/n) \label{EOenergy}\end{equation}
In fact, the states $|EO>$ and $|\alpha_r>$ are not orthogonal.  However, their
overlap (divided by $<EO|EO>$is order $1/n$
(see Appendix \ref{app:overlaps}) and thus the effect of the
non-orthogonality is order $1/n^2$.  Therefore, the above result holds to
order $1/n$.

Comparing the values found in equations (\ref{Oenergy}) and (\ref{EOenergy})
we see
that the symmetric ground state is the true ground state (having the lower
energy) which agrees with known results.  The energy difference per site is
$1/n+O(1/n^2)$.

Considering states of the form
\begin{eqnarray}
|\psi_1>&=&|0>+(a/L)\sum_r|\beta_r>\nonumber \\
|\psi_2>&=&|EO>+[b/(2L)]\sum_r|\alpha_{2r}>\end{eqnarray}
and calculating the ground state energies variationally produces the
same results as those found above to $O(1/n)$ (see Appendix \ref{app:variation}).
Extrapolating the large-n approximation to n=2, and taking into account
the shift of the Hamiltonian by a constant noted in (\ref{Hshift}), the
estimates of the ground state per site to $0^{th}$ and $1^{st}$ order in
$1/n$ are 0 and -1 respectively.  The numerical result for $n=2$ is
-1.401485. \cite{Golinelli}
Thus we see that the inclusion of the terms of first
order in $1/n$ brings us closer to the numerical result.

\subsection{The First Excited States}
\begin{figure}
\epsfxsize=\figwidth
\centerline{\epsffile{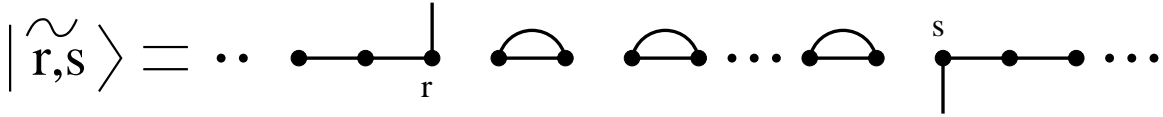}}  
\caption{Soliton-antisoliton state. Note that an uncontracted upper
index is represented by an upwards facing free link while an uncontracted
lower index is represented by a downwards facing free link.}
\label{fig:solitondef}\end{figure}

Since our Hamiltonian counts the number of nearest neighbor bonds to leading
order in $1/n$, the excited states consist of an uncontracted lower index at some
site and an uncontracted upper index at some other site (that is, they have one
fewer nearest neighbor bond).  These uncontracted indices can be though of
as solitons/antisolitons interpolating between two ``ground states''.  We
label such sites as in Figure \ref{fig:solitondef}.
Note that $r$ labels the uncontracted upper index and $s$ labels
the uncontracted lower index.  
We will assume that the indices on
sites $r$ and $s$ are different to eliminate mixing with the singlet
states (in which the indices on sites $r$ and $s$ are contracted).
We define
\begin{equation} x \equiv r-s \end{equation}
Note that $r$ must be even and $s$ must be odd implying that $x$
is odd.
Since the energy of the non-symmetric state is higher than that of
the symmetric state by $1/n$ $(+O(1/n^2))$,
we expect the soliton-antisoliton pair
to experience a linear (in $x$) confining potential like that encountered
in quark confinement.

Let us first consider $H|\widetilde{r,s}>$ for $|s-r|>1$.
We need only consider the effects of $H_{r-1}, H_r, H_{s-1}, H_s$.
The effects of the terms in the Hamiltonian acting on the other links have
already been calculated.

We will first consider the effect of $H_{r-1}$.  We let $i_1$ be the index
contracted between the sites $r-2$ and $r-1$, $i_2$ the index contracted between
sites $r-1$ and $r$, and $i_3$ the free index on site $r$.  Now,
\begin{eqnarray}
H_{r-1}|_{i_1i_2},^{i_2i_3}>&=&-(1/n)[\delta^{i_2}_{i_1}|_{mi_2},^{mi_3}>+
\delta^{i_2}_{i_2}|_{i_1m},^{mi_3}>+\delta^{i_3}_{i_1}|_{mi_2},^{i_2m}>+
\delta^{i_3}_{i_2}|_{i_1m},^{i_2m}>]\nonumber \\
&=&-(1/n)[|_{mi_1},^{mi_3}>+n|_{i_1m},^{mi_3}>+
\delta^{i_3}_{i_1}|_{mi_2},^{i_2m}>+|_{i_1m},^{i_3m}>]\nonumber \\
&=&-(1/n)[(n+2)|_{mi_1},^{mi_3}>+\delta^{i_3}_{i_1}|_{mi_2},^{i_2m}>]
\end{eqnarray}
We recognize the first term as a constant times the original state (with $i_2$
renamed $m$).  The second term represents a double bond between states $r-1$
and $r$ while the delta function serves to create a free index
(equal to the free index in the original state) on site $r-2$.
Thus, from this second term, we obtain the state $|\widetilde{r-2,s}>$.
And so,
\begin{equation}
H_{r-1}|\widetilde{r,s}>=-(1+2/n)|\widetilde{r,s}>-(1/n)|\widetilde{r-2,s}>
\end{equation}
We now consider the effect of $H_r$.  Using the same indices as above
and letting $i_4$ and $i_5$ be the indices representing the double bond
between sites $r+1$ and $r+2$, we calculate
\begin{equation}
H_r|^{i_2i_3},_{i_4i_5}>=-(1/n)[\delta^{i_4}_{i_2}|^{mi_3},_{mi_5}>+
\delta^{i_4}_{i_3}|^{i_2m},_{mi_5}>+\delta^{i_5}_{i_2}|^{mi_3},_{i_4m}>+
\delta^{i_5}_{i_3}|^{i_2m},_{i_4m}>]\end{equation}
Now, in the second and fourth terms above, the index $i_2$ will contract to the
left creating a bond, the index $m$ represents a bond between sites $r$
and $r+1$, the
indices $i_5$ and $i_4$ (respectively) produce a bond between sites $r+1$ and
$r+2$, and the delta function serves to move the soliton (same free index) to site
$r+2$.  Thus from these terms, we obtain the state $|\widetilde{r+2,s}>$.  In the
first and third terms above, the soliton (represented by the free
index $i_3$) remains on site $r$, the index $m$ represents a
bond between sites $r$ and $r+1$, and the indices $i_5$ and $i_4$
(respectively) produce a bond between sites $r-1$ and $r+2$.  Thus, in effect,
we have replaced the bond between sites $r-1$ and $r$ and one of the bonds
between sites $r+1$ and $r+2$ with one bond between sites $r$ and $r+1$ and one
between sites $r-1$ and $r+2$.  Thus, we have reduced the number of nearest
neighbor bonds by one.  This increases the energy by order 1.  Therefore,
we ignore these terms in our analysis.  So,
\begin{equation}H_r|\widetilde{r,s}>=-(2/n)|\widetilde{r+2,s}>\end{equation}
By symmetry,
\begin{eqnarray}
H_s|\widetilde{r,s}>&=&-(1+2/n)|\widetilde{r,s}>-(1/n)|\widetilde{r,s+2}>
\nonumber \\
H_{s-1}|\widetilde{r,s}>&=&-(2/n)|\widetilde{r,s-2}>
\end{eqnarray}
Therefore, neglecting terms with energy higher by an amount of order 1,
\begin{eqnarray}
H|\widetilde{s,r}>&=&-[(L-1)(1+2/n)-(|x|-1)/n]|\widetilde{r,s}>
-(2/n)(|\widetilde{r+2,s}>+|\widetilde{r,s-2}>)\nonumber \\
&&\ -(1/n)(|\widetilde{r-2,s}>+|\widetilde{r,s+2}>)
\ \ \hbox{(for $|r-s|>1$)}\end{eqnarray}

We now consider the action of the Hamiltonian on the state
$|\widetilde{r,r+1}>$ (that is, for $x=1$).
The action on $|\widetilde{r,r-1}>$ ($x=-1$)
will follow by symmetry.  We only need to calculate
the effect of $H_r$; the rest are known.  Assume the index $i_1$ is contracted
to the left, $i_4$ is contracted to the right, and $i_2$, $i_3$ are free
and consider
\begin{equation}
H_r|_{i_1i_2},^{i_3i_4}>=-(1/n)[\delta^{i_3}_{i_1}|_{mi_2},^{mi_4}>
+\delta^{i_3}_{i_2}|_{i_1m},^{mi_4}>+\delta^{i_4}_{i_1}|_{mi_2},^{i_3m}>+
\delta^{i_4}_{i_2}|_{i_1m},^{i_3m}>]\end{equation}
Now, the first term results in the state $|r,r-1>$.  The second term results
in a delta function times the symmetric ground state.  The third term
results in a state with a soliton on site r, an antisoliton on site r+1 and
bonds between sites r and r+1 as well as r-1 and r+2 (this state has one
fewer nearest neighbor bonds).  The fourth term results in the state
$|r+2,r+1>$.  Thus, ignoring states with higher energy of order 1
and assuming that $i_2\ne i_3$,
\begin{equation}H_r|\widetilde{r,r+1}>=-(1/n)(|\widetilde{r,r-1}>
+|\widetilde{r+2,r+1}>)
\end{equation}
and similarly
\begin{equation}H_r|\widetilde{r,r-1}>=-(1/n)(|\widetilde{r-2,r-1}>
+|\widetilde{r,r+1}>)
\end{equation}

At first glance, it seems as though our Hamiltonian is non-hermitian.
However, we can see that this is simply due to the normalization of
the $|\widetilde{r,s}>$ states.
The normalization for sites $r+1$ through $s-1$ is the same as that for
an $|EO>$ state of length $|x|-1$.  We see from Appendix \ref{app:overlaps} that this
normalization is $[2n(n+1)]^{(|x|-1)/2}$.
The normalization for the rest of the
sites is the same as that for an $|0>$ state of length $L-|x|+1$ (where $L$ is
the total number of sites in the state $|r,s>$).  It is shown in Appendix
\ref{app:overlaps}
that this normalization is $((n+1)^{L-|x|+1}-1)/n$.
Thus the total normalization is
\begin{equation}
<\widetilde{r,s}|\widetilde{r,s}>=[2n(n+1)]^{(|x|-1)/2}[((n+1)^{L-|x|+1}-1)/n]
\end{equation}
In the large L and large n limits (taking L large first as usual), this
reduces to $2^{(|x|-1)/2}n^{L-1}$.  Thus we define the ``properly''
normalized states
\begin{equation}
|r,s>=2^{-(|x|-1)/4}n^{(1-L)/2}|\widetilde{r,s}>
\label{rsdef}
\end{equation}
With this normalization, the Hamiltonian becomes
\begin{eqnarray}
(H+(L-1)(1+2/n))|r,s>&=&[(|x|-1)/n]|r,s>-(\sqrt2/n)(|r+2,s>+|r,s-2>\nonumber \\
&&+|r-2,s>+|r,s+2>)\hbox{ for } |x|>1 \nonumber \\
(H+(L-1)(1+2/n))|r,r+1>&=&-(1/n)(|r,r-1>+|r+2,r+1>) \nonumber \\
(H+(L-1)(1+2/n))|r,r-1>&=&-(1/n)(|r-2,r-1>+|r,r+1>)
\label{Hunmodifieddef}
\end{eqnarray}

Now, we expect the first excited states to be translationally invariant.
Therefore we define the translationally invariant states
\begin{equation}|x>=(1/\sqrt{2L})\sum_r |2r,2r+x>\end{equation}
where the multiplicative constant ensures that $<x|x>=1$. (This follows
from the normalization of $|r,s>$ since it is shown
in Appendix \ref{app:solitonoverlaps} that $<\widetilde{r',s'}|\widetilde{r,s}>$ is of order $n^{L-2}$
or smaller if $r' \ne r$ and/or $s' \ne s$.)

Therefore, for $|x|>1$,
\begin{eqnarray}
(H+(L-1)(1+2/n))|x>&=&(H+(L-1)(1+2/n))[(1/\sqrt{2L})\sum_r|2r,2r+x>]\nonumber \\
&=&(1/\sqrt{2L})\sum_r[((|x|-1)/n)|2r,2r+x>
-(\sqrt2/n)(|2r+2,2r+x> \nonumber \\
&&\ \ +|2r,2r+x-2>+|2r-2,2r+x>+|2r,2r+x+2>)]
\nonumber \\
&=&[(|x|-1)/n]|x>-(2\sqrt2 /n)(|x-2>+|x+2>)
\end{eqnarray}
And for $x=\pm 1$
\begin{eqnarray}
(H+(L-1)(1+2/n))|1>&=&-(2/n)|-1>-(2\sqrt2 /n)|3>\nonumber \\
(H+(L-1)(1+2/n))|-1>&=&-(2/n)|1>-(2\sqrt2/n)|-3>
\end{eqnarray}

Although the assumption will prove to be a poor one in future calculations,
if we assume that our wave function varies slowly, we can change
our Hamiltonian into continuous form by Taylor expanding the states involved to 
obtain
\begin{equation}
|x\pm 2>\approx |x>\pm 2(d/{dx})|x>+2(d^2/{dx^2})|x>
\end{equation}
Therefore,
\begin{equation}|x+2>+|x-2>\approx 2|x>+4(d^2/{dx^2})|x>+...\end{equation}
Making this substitution into the expression for our Hamiltonian and
ignoring the fact that the action of the Hamiltonian is slightly different for
$x=\pm 1$ than it is for $|x|>1$,
\begin{equation}
H=-[(L-1)(1+2/n)-4\sqrt2/n]+(|x|-1)/n-(8\sqrt2/n)d^2/{dx^2}
\label{Hcont}
\end{equation}
This is the Hamiltonian for a particle in a linear potential with $x$ corresponding
to the difference coordinate of the soliton-antisoliton pair.  Thus, in this
approximation, the soliton-antisoliton pair experiences a linear confinement
potential similar to that encountered in quark confinement.  If the pair
becomes too separated, it splits into two pairs much like the quark-antiquark
pair comprising a meson (this will be discussed in more detail later).  The
exact Hamiltonian (in the large-$n$ limit) is of this form but the kinetic energy
term is not equal to $p^2$ as the above approximation would suggest.

We now define a general state
\begin{equation}|\psi>=\sum_{x\hbox{ odd}}\psi(x)|x>\end{equation}
We then seek the lowest energy eigenstate of this form.  This will be the first
excited state.  Thus, we wish to find the function $\psi(x)$,
defined on the odd integers such that
\begin{eqnarray}
(H-(E_0+1+2/n))|\psi>&=&\delta E|\psi>\nonumber \\
\Rightarrow \sum_{x odd}\psi(x)(H-(E_0+1+2/n))|x>&=&\delta E\sum_{x odd}\psi(x)|x>
\label{Eprimedef} \end{eqnarray}
where $E_0=-L(1+2/n)$ is the groundstate energy.  Thus
\begin{eqnarray}
-(2/n)[\psi(-1)|1>+\psi(1)|-1>]
-(2\sqrt2 /n)[\psi(1)|3>+\psi(-1)|-&3&>]
\nonumber \\
+\sum_{x odd, |x|>1}\psi(x)[((|x|-1)/n)|x>-(2\sqrt2 /n)(|x+2>+|x-2>)]&=&
\delta E\sum_{x odd}\psi(x)|x>\end{eqnarray}
We now take the overlap of both sides with the state $<x'|$.  We have from
the calculations of Appendix \ref{app:solitonoverlaps} that the overlap $<x'|x>$ is order $n^{L-1}$
if $x'=x$ and is of smaller order in $n$ if $x'\ne x$.  Thus, from the
overlap with $|1>$ we obtain, to order $1/n$,
\begin{equation}
-(2/n)\psi(-1)-(2\sqrt2/n)\psi(3)=\delta E\psi(1)
\Rightarrow \psi(3)=-(2\psi(-1)+n\delta E\psi(1))/(2\sqrt2)
\label{recursion1}\end{equation}
And for  $|x|>1$ we obtain, to order $1/n$,
\begin{eqnarray}
-(2\sqrt2/n)[\psi(x-2)+\psi(x+2)]+((|x|-1)/n)\psi(x)&=&\delta E\psi(x)\nonumber \\
\Rightarrow \psi(x+2)&=&[(|x|-n\delta E-1)/(2\sqrt2)]\psi(x)-\psi(x-2)
\label{recursion2}\end{eqnarray}
Now, since the Hamiltonian is an even function of $x$, we expect
$\psi(x)$ to be either an even or an odd function of $x$.  We can easily prove
that the first excited state is even by a contradiction argument inspired by
Feynman.  Assume that the first
excited state is odd (that is, $\psi(x)$ is an odd function of $x$).  Now define
\begin{eqnarray}
g(x)&=&-\psi(x), x<0\nonumber \\
&=&\psi(x), x>0\end{eqnarray}
which is obviously an even function of $x$.  Now, the two
states defined by these functions obviously have the same normalization.
Thus, we can ignore this normalization when comparing the energies
of the two states.  The energy of the state $|\psi>$ relative to 
$E_0+1+2/n$ is
\begin{eqnarray}
E_{\psi}&=&-(4/n)\psi(-1)\psi(1)+
\sum_{x odd, |x|>1}[((|x|-1)/n)\psi(x)^2-(2\sqrt2/n)(\psi(x)\psi(x-2)+\psi(x)\psi(x+2))]\nonumber \\
&=&(4/n)\psi(1)^2+2\sum_{x odd, x>1}[((|x|-1)/n)\psi(x)^2-(2\sqrt2/n)(\psi(x)\psi(x-2)+\psi(x)\psi(x+2))]
\end{eqnarray}
while the energy of the state defined by $g(x)$ is
\begin{eqnarray}
E_g&=&-(4/n)g(-1)g(1)+
\sum_{x odd, |x|>1}[((|x|-1)/n)g(x)^2-(2\sqrt2/n)(g(x)g(x-2)+g(x)g(x+2))]
\nonumber \\
&=&-(4/n)\psi(1)^2+2\sum_{x odd, x>1}[((|x|-1)/n)\psi(x)^2-(2\sqrt2/n)(\psi(x)\psi(x-2)+\psi(x)\psi(x+2))]
\nonumber \\
&=&E_{\psi}-(8/n)\psi(1)^2\end{eqnarray}
Thus the function defined by $g(x)$ has a lower (or equal if $\psi(1)=0$)
energy than the state $|\psi>$.  This contradicts the claim
that $|\psi>$ is the first excited state.
Thus the first excited state is an
even function of $x$.

\begin{figure}[htbp]
\epsfxsize=\figwidth
\centerline{\epsffile{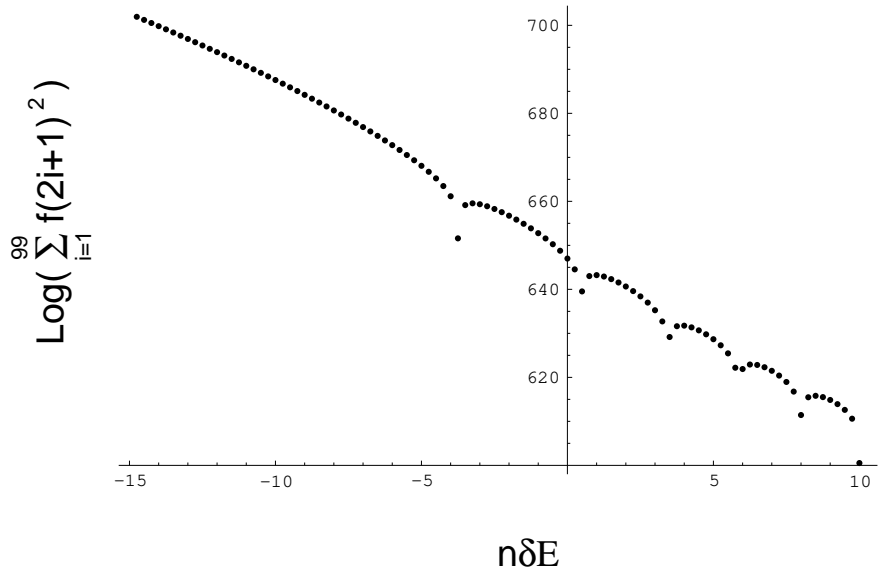}}  
\caption{$Log$ of partial normalization sum versus $n\delta E$.  The $log$ of
the sum is plotted to allow the wide range to be displayed more
clearly.}
\label{fig:elots}\end{figure}

Therefore, to find the first excited state, we can set $\psi(1)=1$
and need only determine
$\psi(x)$ for positive $x$.  Since $\psi$ is even, $\psi(-1)=\psi(1)=1$
and so (\ref{recursion1}) becomes
\begin{equation}\psi(3)=-(n\delta E+2)/{2\sqrt2}\label{recursion3}\end{equation}
Equation \ref{recursion3} gives $\psi(3)$ in terms of $\psi(1)=1$ and $n\delta E$.
Equation \ref{recursion2} is a recursion relation defining the $\psi(x)$ for $x$ odd.
Now, in order for the state $|\psi>$ to be normalizable, we need the sum
$\sum_{z=0}^{\infty}\psi(2z+1)^2$
to converge.  We expect that this will only occur for discrete values of
$n\delta E$.  Thus, using the above recursion relation, we plot the value
of a partial sum of the above series for various values of $n\delta E$.
This plot is shown is Figure \ref{fig:elots}.

\begin{figure}
\epsfxsize=\figwidth
\centerline{\epsffile{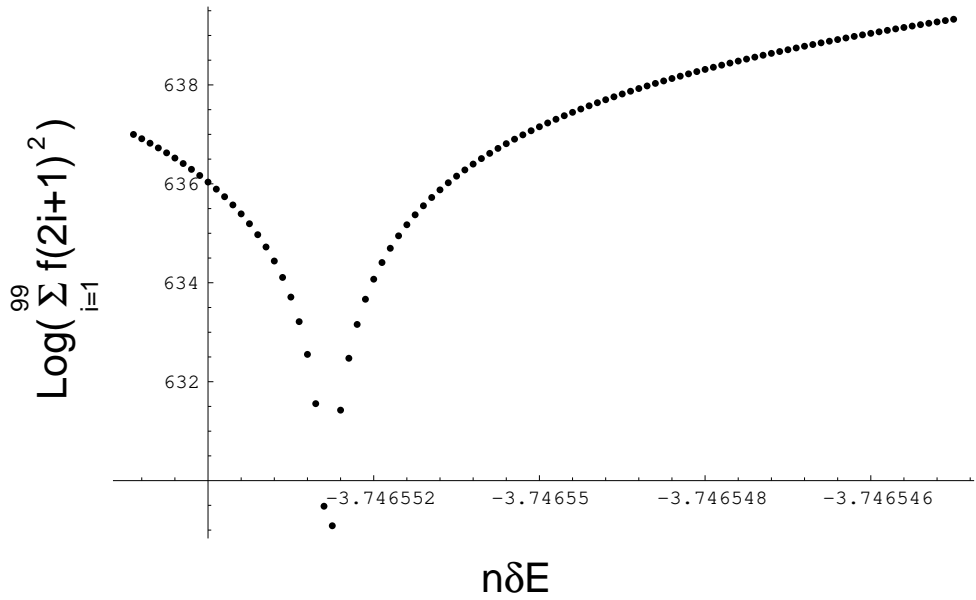}}  
\caption{A blow up of Figure \protect{\ref{fig:elots}} near the first minimum.}
\label{fig:emin}\end{figure}

\begin{figure}
\epsfxsize=\figwidth
\centerline{\epsffile{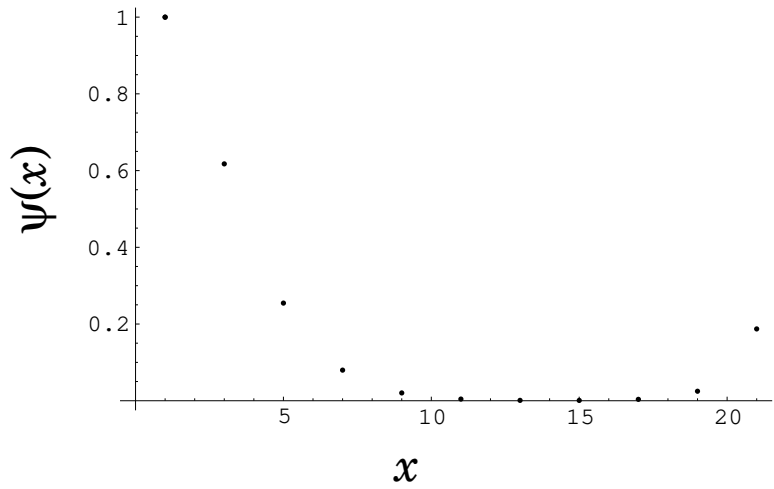}}  
\caption{Ground state wave function.  Since the wave function is even,
it is only plotted for positive $x$.  Note that the wave function starts
to diverge at about $x=19$.  This is due to the fact that the eigenvalue found
is merely a numerical approximation to the true eigenvalue.}
\label{fig:psi0}\end{figure}

We find that the first excited state has $n\delta E \approx -3.747$ (see
Figure \ref{fig:emin}).  The wave function for this energy is shown in
Figure \ref{fig:psi0}.
Thus
the energy gap of the first excited state to order $1/n$ is
\begin{equation}
\Delta =E_1-E_0=1+2/n+\delta E\approx 1+2/n-3.747/n=1-1.747/n
\end{equation}
Thus, for n=2, $\Delta \approx 0.1265$.  Numerically and
experimentally, this value is
found to be 0.41050\cite{White}.  Since we obtain $\Delta =1$ to
leading order, and $\Delta \approx 0.1265$ to order $1/n$ it seems
plausible that were one to include higher orders, the calculated value
of $\Delta$ might approach 0.41050 in an oscillatory manner.

Using our numerical results to order $1/n$ we can also calculate
how many bound
states exist.  We assume that a necessary and sufficient condition for
the stability of a given state is that it has less than twice the
energy of the first excited state (otherwise, it would decay into two
or more states of lower energy).  Thus, the $m^{th}$ excited state
will be stable if and only if
\begin{eqnarray}
1+2/n+(n\delta E)_m/n&<&2(1-1.747/n) \nonumber \\
\Rightarrow (n\delta E)_m&<&n-5.494
\end{eqnarray}
Now, we find that the second excited state is a parity odd state with
$(n\delta E)_2 \approx 0.467$.  Thus, the second excited state
is stable if and only if
\begin{equation} 0.467<n-5.494 \Rightarrow n>5.96 \end{equation}
Thus, for $n=2$ there exists only one stable magnon in agreement
with $\sigma$-model, numerical and experimental results.

\section{Single Modified Link}
We now consider the case in which one of the links in our chain has a coupling
which may differ from that of the other links.
In the following discussion, we will find it useful to use the coordinates
\begin{eqnarray}
x&\equiv&s-r \nonumber \\
y&\equiv&s+r-1
\end{eqnarray}
rather than $r$ and $s$ to label our magnon states.
The fact that $r$ must be even and $s$ must be odd implies that either
$x=4p+1,y=4q$ or $x=4p+3,y=4q+2$ where $p$ and $q$ are both integers.
Thus although $x$ must be odd and $y$ must be even, not all combinations
of odd $x$ and even $y$ are allowed.

We define the coupling between
sites 0 and 1 to be $J'$ and between all the other sites to be $J$.  As in the
previous calculations, we will choose our units of energy so that $J=1$.  We
assume that $J'-1$ is $O(1/n)$.  This implies that the ground state will
remain unchanged to leading order in $1/n$ and that its energy will be 
(to order $1/n$)
\begin{equation}
E_0=-(L+J'-1)(1+2/n)
\end{equation}
Note that this redefinition of $E_0$ corresponds with our earlier definition
in the case $J'=1$.
We now define $H_0$ to be the Hamiltonian without the modified link with
the new generalized $E_0$ substituted for the old (that is, we define
$H_0-E_0-(1+2/n)$ to be the right sides of (\ref{Hunmodifieddef})).

\begin{figure}
\epsfxsize=\statewidth
\centerline{\epsffile{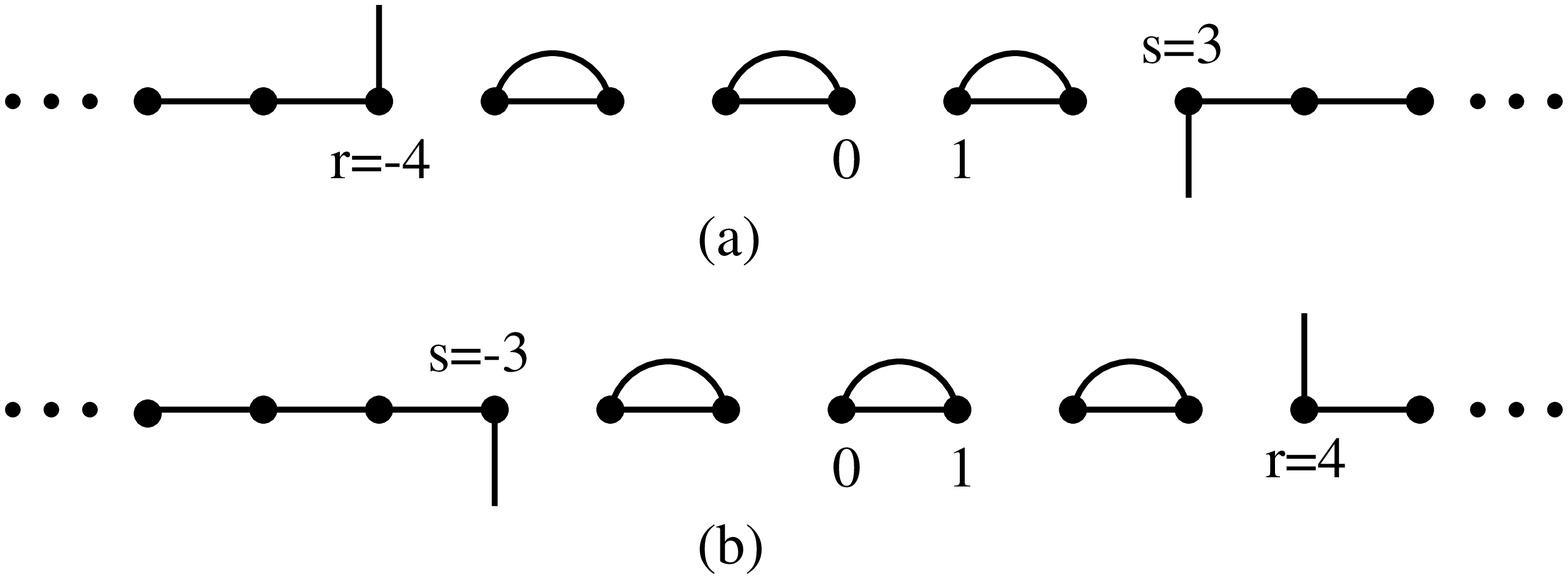}}  
\caption{(a) Typical configuration with $r \le 0$, $s \ge 1$ ($\Rightarrow x \ge 1$).
There is no bond on the modified link 01.
(b) Typical configuration with $r \ge 2$, $s \le -1$ ($\Rightarrow x \le -1$).
There is a double bond on the modified link 01.}
\label{fig:potential}\end{figure}

We can write the true
Hamiltonian (with the modified link) as $H_0$ plus terms correcting
for the modified link.  That is, we can write the true Hamiltonian $H$ as
\begin{equation}H=H_0+\lambda H_k+\lambda V\end{equation}
where $\lambda=1-J'=O(1/n)$ and
\begin{eqnarray}
H_k|r,s>&=&(\sqrt2/n)|r+2,s>\hbox{ for }r=0,\  s\ne 1 \nonumber \\
&=&(\sqrt2/n)|r,s-2>\hbox{ for }s=1,\ r\ne 0\nonumber \\
&=&(1/n)[|r+2,s>+|r,s-2>]\hbox{ for } r=0,\ s=1\nonumber \\
&=& 0 \ \ \hbox{otherwise}\nonumber \\
V|r,s>&=&-(1+2/n)|r,s>\hbox{, for }r\le 0,s\ge 1\nonumber \\
&=&+(1+2/n)|r,s>\hbox{, for }r\ge 2,s\le -1 \nonumber \\
&=&0 \ \ \hbox{otherwise}
\label{VHkrsdef}
\end{eqnarray}
Thus $\lambda H_k$ is a modified kinetic energy operator and $\lambda V$ is
a modified potential.  Now, since we are considering the large $n$ limit,
we can neglect $H_k$ and drop the $2/n$ term in $V$ since these are lower
order in $n$.  [Recall that $\lambda$ is $O(1/n)$.]
Thus, if we now change coordinates to $x$ and $y$, (\ref{VHkrsdef}) becomes
to leading order in $1/n$
\begin{eqnarray}
H_k|x,y>&=&0 \nonumber \\
V|x,y>&=&-|x,y>\hbox{, for }x\ge 1,|y|\le x-1 \nonumber \\
&=&+|x,y>\hbox{, for }x\le -3,|y|\le -(x+3)\nonumber \\
&=&0 \ \ \hbox{otherwise}
\label{VHkdef}
\end{eqnarray}
\begin{figure}
\epsfxsize=\figwidth
\centerline{\epsffile{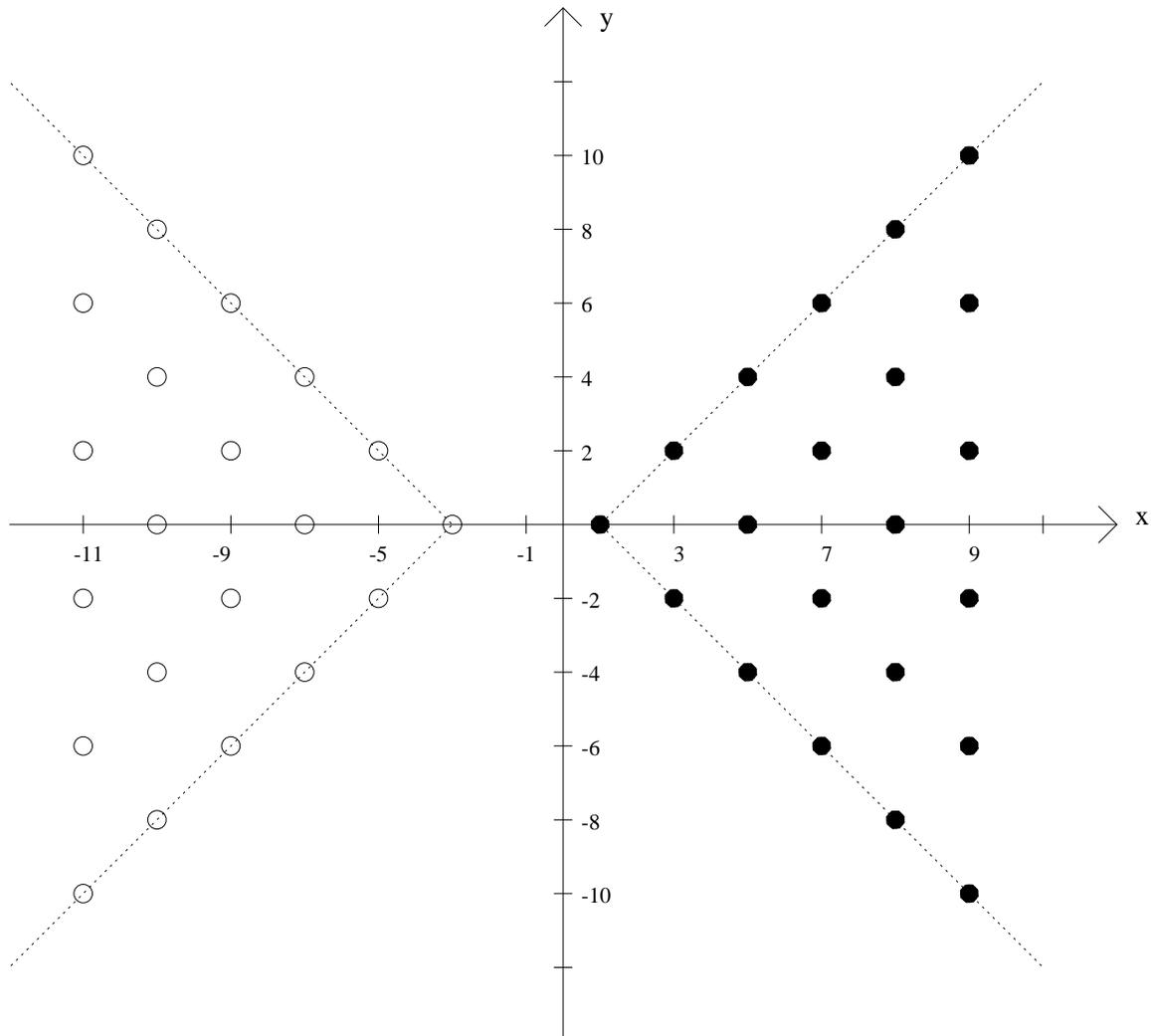}}  
\caption{Modified Potential $V$.  An open circle denotes a coordinate
at which $V$ has the value +1 and a closed circle denotes a coordinate
at which $V$ has the value -1}
\label{fig:vplot}\end{figure}
Thus, the modified link not only breaks translational symmetry, it
also breaks parity symmetry (reflection about a site) since $V$
is not symmetric with respect to parity.
Our potential arises from the fact we have defined our energy such
that zero energy corresponds to the modified link being outside
the soliton-antisoliton pair where there is a single bond between
neighboring sites.  If the modified link is instead between the soliton
and the antisoliton, it is either between two sites not bonded together
(in this case, $V$=-1) or between two sites with a double bond (in this
case $V$=+1).  We see from (\ref{VHkdef}) and Figure \ref{fig:potential}
that if the modified link is between the soliton and antisoliton
(that is, $V$ is non-zero) then, it must lie between two sites
not bonded together if $x\ge 1$ and on a double bond if $x \le -3$.

Our goal is to determine the nature of possible bound states of the
Hamiltonian given above.  Let us define
\begin{equation}
|\psi_0>=\sum_{\hbox{allowed }x,y}\psi_0(x)|x,y>
\end{equation}
to be the
translationally invariant state we found in the case of the unmodified
Hamiltonian, normalized so that
\begin{equation}
\sum_{x \hbox{ odd}}\psi_0(x)^2=1
\label{psi0norm}\end{equation}
We see that in the limit of large $L$ (recall
$L$ is the length of our chain), $|\psi_0>$ has energy
$E_0 + \Delta$ (note the change in the definition of $E_0$ here)
since the state is spread along the entire length of the chain
(that is, it is independent of $y$)
while the modified link occurs only between two sites and thus its effect
becomes negligible.  Therefore, we expect that an eigenstate of the
modified Hamiltonian will be bound (that is, be normalizable) if and only
if it has energy less than $E_0 + \Delta$.  Let $|\phi>$ be such
a bound eigenstate where
\begin{equation}
|\phi>=\sum_{\hbox{allowed }x,y}\phi(x,y)|x,y>
\end{equation}
Since $|\phi>$ is a bound state, we know that $<\phi|\phi><\infty$.
Thus, the energy of the state $|\phi>$ is
\begin{eqnarray}
E_{\phi}&=&<\phi|H_0|\phi>/<\phi|\phi>+\lambda<\phi|V|\phi>/<\phi|\phi>
\nonumber \\
&\ge&E_0+\Delta+\lambda<\phi|V|\phi>/<\phi|\phi>
\label{Ephi}
\end{eqnarray}
where the inequality is due to the fact that, by definition, $\Delta$
is the lowest energy eigenstate of $H_0-E_0$ (and therefore
$E_0+\Delta$ is the lowest energy eigenstate of $H_0$).
Now, we know that $\phi(x,y)$ must decay rapidly at large $|x|$ due to the
linear potential present in the Hamiltonian.
We see from (\ref{VHkdef}) and Figure \ref{fig:vplot}
that for small $x$, the potential $V$ is only non-zero for small $y$.
Since our perturbation is small (order $\lambda$), we expect
that for small $x$ and $y$, $\phi\approx\psi$ (that is, the corrections
are order $\lambda$ and thus can be ignored to leading order).
Therefore (\ref{Ephi}) becomes, to leading order in $\lambda$,
\begin{equation}
E_{\phi}\ge E_0 + \Delta+\lambda<\psi_0|V|\psi_0>/<\phi|\phi>
\end{equation}
The fact that $\psi_0(x)$ is independent of $y$ makes
$<\psi_0|V|\psi_0>$ particularly easy to calculate.
Since $\psi_0(x)$ is an even function of $x$, both signs of $x$
are weighted evenly in the sum $<\psi_0|V|\psi_0>$.  Also, the
fact that $\psi_0(x)$ is independent of $y$ means that the weighting
is the same (for a given value of $|x|$) no matter where the modified
link is.
Now, for a given $|x|$, there is always one more link
with no bond than with a double bond.  This can be seen easily
in Figure \ref{fig:potential}.  Thus, for each $|x|$, the sum
over $+x$ and $-x$ and the respective allowed values of $y$ in
$<\psi_0|V|\psi_0>$ equals -1 since $V$ takes on the value -1 one
more time than it takes on the value +1.  Thus,
\begin{equation}
<\psi_0|V|\psi_0>=-\sum_{x\ge 1\hbox{ odd}}\psi_0(x)^2=-1/2
\end{equation}
since this sum is equal to half the normalization sum in (\ref{psi0norm}).
Therefore,
\begin{equation}
E_{\phi}\ge E_0 + \Delta-\lambda/(2<\phi|\phi>)
\end{equation}
Since $<\phi|\phi>>0$, if $\lambda<0$ then $E_{\phi}\ge E_0+\Delta$
which is a contradiction since this implies $|\phi>$ is not a bound
state.  Thus, in the small $\lambda$ limit,
it is only possible for a bound state to occur
if $\lambda>0$ (that is, if $J'<J$).
However, we do not yet know that a bound state
{\it does} exist in this case.  To see this, we will use a variational
argument.

We choose as our variational state the normalized wavefunction,
\begin{eqnarray}
|\psi>&=&\sum_{\hbox{allowed }x,y}\psi(x,y)|x,y> \nonumber \\
\hbox{where }\psi(x,y)&=&A\psi_0(x)e^{-\epsilon |y|}
\hbox{ and } A \rightarrow \sqrt{4\epsilon}
\hbox{ as }\epsilon \rightarrow 0
\label{varwf}
\end{eqnarray}
To see how the normalization constant $A$ (chosen so that $<\psi|\psi>=1$)
was obtained, recall that either
$x=4p+1, y=4q$ or $x=4p+3, y=4q+2$ where $p$ and $q$ are integers.
Thus,
\begin{eqnarray}
\sum_{\hbox{allowed }y}&&(e^{-\epsilon|y|})^2 \nonumber \\
&=&\sum_{q \epsilon Z}e^{-8\epsilon|q|}
=2\sum_{q=0}^{\infty}e^{-8\epsilon q}-1=2/(1-e^{-8\epsilon})-1
=(1+e^{-8\epsilon})/(1-e^{-8\epsilon}) \approx 1/(4\epsilon) \nonumber \\
OR \ \ \  &=&2e^{-4\epsilon}\sum_{q=0}^{\infty}e^{-8\epsilon|q|}
=2e^{-4\epsilon}/(1-e^{-8\epsilon}) \approx 1/(4\epsilon)
\end{eqnarray}
which gives us our normalization constant A in the small $\epsilon$ limit.
We choose the trial wave function given above since it
reduces to the translationally
invariant wave function in the limit $\epsilon \rightarrow 0$ (we
expect $\epsilon \rightarrow 0$ as $\lambda \rightarrow 0$).
We will minimize the energy of this trial wave function
with respect to $\epsilon$.

In terms of the coordinates $x$ and $y$ and using the same
normalization for the $|x,y>$ states as was used for the
$|r,s>$ states, $H_0$ is given by
\begin{eqnarray}
(H_0-(E_0+1+2/n))|x,y>&=&((|x|-1)/n)|x,y>-(\sqrt2/n)(|x+2,y+2>+|x+2,y-2>
\nonumber \\
&&+|x-2,y+2>+|x-2,y-2>) \hbox{ for } |x|>1 \nonumber \\
(H_0-(E_0+1+2/n))|1,y>&=&-(\sqrt2/n)(|3,y+2>+|3,y-2>) \nonumber \\
&&-(1/n)(|-1,y+2>+|-1,y-2>) \nonumber \\
(H_0-(E_0+1+2/n))|-1,y>&=&-(\sqrt2/n)(|-3,y+2>+|-3,y-2>) \nonumber \\
&&-(1/n)(|1,y+2>+|1,y-2>)
\label{H0xydef}\end{eqnarray}
We would like to separate $H_0$ into its $x$ and $y$ components.  One possible
method for doing this would be to introduce lattice derivatives.  However,
a problem arises when we attempt this.  We would like to add and subtract
the states $|x+2,y>$ and $|x-2,y>$ from the right hand side of
(\ref{H0xydef}).  However, given that the state $|x,y>$ exists,
neither of the states $|x+2,y>$ or $|x-2,y>$ can exist (this
follows from the permissible values of $x$ and $y$).  However,
the states $|-(x+2),y>$ and $|-(x-2),y>$ do exist.  Although adding
and subtracting these states from the right hand side of (\ref{H0xydef})
may seem counterintuitive, the fact that $\psi(x,y)$ is an even
function of $x$ allows us to effectively separate $H_0$ into its
$x$ and $y$ components in this way.  Thus, we rewrite
the $|x|>1$ portion of (\ref{H0xydef}) as
\begin{eqnarray}
(H_0-(E_0+1+2/n))|x,y>&=&((|x|-1)/n)|x,y>-(\sqrt2/n)(|x+2,y+2>+|x+2,y-2>
-2|-(x+2),y>\nonumber \\
&&+|x-2,y+2>+|x-2,y-2>)-2|-(x-2),y>) \nonumber \\
&&-(2\sqrt2/n)(|-(x+2),y>+|-(x-2),y>-2|x,y>)-(4\sqrt2/n)|x,y> \nonumber \\
&=&((|x|-1)/n-4\sqrt2/n)|x,y>-(\sqrt2/n)(\Delta_y|x+2,y>+\Delta_y|x-2,y>)
\nonumber \\
&&-(2\sqrt2/n)\Delta_x|x,y>
\label{H0def1}\end{eqnarray}
Where
\begin{eqnarray}
\Delta_x|x,y>&=&|-(x+2),y>+|-(x-2),y>-2|x,y> \nonumber \\
\Delta_y|x,y>&=&|x,y+2>+|x,y-2>-2|-x,y>
\end{eqnarray}
are our lattice second derivative operators.  Similarly,
\begin{eqnarray}
(H_0-(E_0+1+2/n))|1,y>&=&-(\sqrt2/n)\Delta_y|3,y>-(1/n)\Delta_y|-1,y>
-(2\sqrt2/n)|-3,y>-(2/n)|1,y> \nonumber \\
(H_0-(E_0+1+2/n))|-1,y>&=&-(\sqrt2/n)\Delta_y|-3,y>-(1/n)\Delta_y|1,y>
-(2\sqrt2/n)|3,y>-(2/n)|-1,y>
\label{H0def2}\end{eqnarray}

We can now roughly separate $H_0$ into its $x$ and $y$ components.
Define $H_0^y$ as the sum of the terms in (\ref{H0def1}) 
and (\ref{H0def2}) involving
$\Delta_y$ and $H_0^x$ as the sum of the remaining terms.
Since $\psi_0(x)$ is an even function of $x$ and is independent of $y$,
$\Delta_y|\psi_0>=0$.  Thus $H_0^y|\psi_0>=0$ and so
$H_0^x$ is precisely the Hamiltonian we encountered in the
case $J'=J$ up to a constant relating to the redefinition of $E_0$.

We now calculate the expectation energy of our trial wavefunction.
\begin{eqnarray}
E_{\epsilon}&=&<\psi|H|\psi> \nonumber \\
&=&E_0+\Delta +<\psi|H_0^y|\psi>+\lambda<\psi|V|\psi>
\end{eqnarray}
Now, although $\Delta_y$ is defined as an operator on the state $|x,y>$,
through a change of variables, we can treat it as an operator on the
coefficients $\psi(x,y)$.
For $y>0$ (note that y is always even),
\begin{eqnarray}
\Delta_y\psi(x,y)&=&\sqrt{4\epsilon}\psi_0(x)[e^{-\epsilon(y+2)}
+e^{-\epsilon(y-2)}-2e^{-\epsilon y}] \nonumber \\
&=&\psi(x,y)[e^{-2\epsilon}+e^{2\epsilon}-2] \nonumber \\
&=&\psi(x,y)[4\epsilon^2+O(\epsilon^4)]
\end{eqnarray}
where we have used the fact that $\psi_0(x)$ is an even function.
Similarly, for $y<0$,
\begin{equation}
\Delta_y\psi(x,y)=\psi(x,y)[4\epsilon^2+O(\epsilon^4)]
\end{equation}
Now, for $y=0$,
\begin{eqnarray}
\Delta_y\psi(x,0)&=&\sqrt{4\epsilon}\psi_0(x)[2e^{-2\epsilon}-2] \nonumber \\
&=&\psi(x,0)[-4\epsilon+4\epsilon^2+O(\epsilon^3)]
\label{deltay} \end{eqnarray}
Notice that the case $y=0$ agrees with the case $y \ne 0$ in the 
$\epsilon^2$ term.
Therefore, dropping terms of order $\epsilon^3$ and higher and
using the fact that $\psi(x,y)$ is an even function of $x$
we see that
\begin{eqnarray}
<\psi|H_0^y|\psi>&=&
-(\sqrt2/n)4\epsilon^2\sum_{\hbox{allowed }|x|>1,y}[\psi(x,y)\psi(-(x-2),y)+\psi(x,y)\psi(-(x+2),y)]
\nonumber \\
&&-(\sqrt2/n)(-4\epsilon)\sum_{\hbox{allowed }|x|>1}[\psi(x,0)\psi(-(x-2),0)+\psi(x,0)\psi(-(x+2),0)]
\nonumber \\
&&-4\epsilon^2\sum_{\hbox{allowed }y}[(\sqrt2/n)\psi(1,y)\psi(-3,y)+(1/n)\psi(1,y)\psi(1,y)]
\nonumber \\
&&+4\epsilon[(\sqrt2/n)\psi(1,0)\psi(-3,0)+(1/n)\psi(1,0)\psi(1,0)]
\nonumber \\
&&-4\epsilon^2\sum_{\hbox{allowed }y}[(\sqrt2/n)\psi(-1,y)\psi(3,y)+(1/n)\psi(-1,y)\psi(-1,y)]
\label{H0ycalc0} \end{eqnarray}
Now, we can use the fact that $\psi(-x,y)=\psi(x,y)$ to convert sums over
all $x$ to (two times) sums over positive $x$.  Also, the $x=\pm 1$ cases
missing in the first two sums above, are present in the third and fourth
sums.  Finally, in the second sum of (\ref{H0ycalc0}), 
 we must remember that only certain values of $x$
 are allowed if $y=0$.  The allowed values are ...,-7,-3,1,5,9,....
 Thus, using the fact that $\psi_0(x)$ is an even function of $x$, we
 can convert a sum over these allowed values into a sum over all odd
 positive $x$.  Thus, we may simplify the above expression to:
\begin{eqnarray}
<\psi|H_0^y|\psi>&=&
-(\sqrt2/n)4\epsilon^2[4\sum_{\hbox{allowed }x\ge1,y}\psi(x,y)\psi(x+2,y)]
\nonumber \\
&&+(\sqrt2/n)(4\epsilon)[2\sum_{\hbox{odd }x\ge1}\psi(x,0)\psi(x+2,0)]
\nonumber \\
&&-(1/n)4\epsilon^2[2 \sum_{\hbox{allowed }y} \psi(1,y)\psi(1,y)]
\nonumber \\
&&+(4\epsilon/n)\psi(1,0)\psi (1,0)
\label{H0ycalc}
\end{eqnarray}
Now, a factor of $\epsilon$ occurs
from the normalization
of $\psi$ which is canceled when a sum over $y$ is performed.  Thus, all the
terms in the above end up being $O(\epsilon^2)$.
 Thus, the above becomes
\begin{eqnarray}
<\psi|H_0^y|\psi>&=&
-(\sqrt2/n)16\epsilon^2\sum_{\hbox{odd }x\ge1}\psi_0(x)\psi_0(x+2)
+(\sqrt2/n)8\epsilon\sum_{\hbox{odd }x\ge1}4\epsilon\psi_0(x)\psi_0(x+2)
\nonumber \\
&&-(8\epsilon^2/n)\psi_0(1)^2
+(16\epsilon^2/n)\psi_0(1)^2 \nonumber \\
&=&(\epsilon^2/n)[16\sqrt2 \theta +8\psi_0(1)^2]
\end{eqnarray}
where
\begin{equation}
\theta = \sum_{\hbox{odd }x\ge 1}\psi_0(x)\psi_0(x+2) >0
\end{equation}
It is shown in Appendix \ref{app:positivity}
that $\psi_0$ is everywhere positive and so we know $\theta$ is positive.

We now consider the term $V$,
Using the same reasoning as when we were calculating $<\phi|V|\phi>$
(that is, for small $x$, $y$ and $\epsilon$,
$\psi(x,y) \approx A\psi_0(x)$), we see that
\begin{equation}
<\psi|V|\psi>\approx A^2<\psi_0|V|\psi_0>=(4\epsilon)(-1/2)=-2\epsilon
\end{equation}

Therefore, the expectation energy of our trial wave function is
\begin{eqnarray}
E_\epsilon&=&E_0+\Delta + 8\kappa\epsilon^2/n
-2\lambda\epsilon \nonumber \\
\hbox{Where }\kappa&=&2\sqrt2\theta+\psi_0(1)^2>0
\end{eqnarray}
In order to minimize this quantity with respect to $\epsilon$, we
set $dE/d\epsilon=0$ to obtain
\begin{eqnarray}
0&=&16\kappa\epsilon/n-2\lambda \nonumber \\
\Rightarrow \epsilon&=&n\lambda/(8\kappa)
\label{epsilon}
\end{eqnarray}
which gives an energy of (to leading order in $n$)
\begin{equation}
E_{min}=E_0+\Delta -n\lambda^2/(8\kappa) < E_0+\Delta
\label{Emin}
\end{equation}
Thus, we know that the true magnon state of the modified link Hamiltonian
has an energy which is less than $E_0+\Delta$ and thus is bound.  Note
that our minimizing value of $\epsilon$ is positive which is necessary
for our trial wave function to be normalizable.  Numerically, $\theta$ is
calculated to be approximately 0.279 and thus $\kappa$ is approximately
1.134.  Therefore, for $n=2$ and $\lambda>0$,
\begin{equation}
E_{min} \approx E_0+\Delta -0.221\lambda^2
\label{benum} \end{equation}

So far we have presented the wave-function of (\ref{varwf}) as merely a 
variational one so that the energy $E_{min}$ is just an upper bound.  
However, we expect the variational wave-function to become sufficiently 
accurate that the actual boundstate energy is given by (\ref{Emin}) in the 
limit $\lambda \to 0^+$.  This follows from the following observations.  
First of all, $\psi (x,y)$ is an exact eigenfunction in the translationally 
invariant case, $\lambda =0$, for any $\epsilon$.  Secondly, even for 
$\lambda \neq 0$, $\psi (x,y)$ satifies exactly the Schroedinger equation
outside the dotted lines in Figure \ref{fig:vplot}.
For any fixed value of $x$ and 
sufficiently small $\lambda$ [and hence $\epsilon$ as determined by
(\ref{epsilon})] 
$\psi (x,y)$ is nearly constant as a function of $y$ between the dotted 
lines of Figure \ref{fig:vplot}.
The true boundstate wave function must also have this 
property in order for $<\psi |H_0^y|\psi >$ to be small.  Thus the actual 
value of the wave-function in this region is not important as long as it 
is nearly constant and joins smoothly with the wave-function outside the 
dotted lines, properties enjoyed by $\psi (x,y)$.  At $|x|>>1$, $\psi (x,y)$ 
goes to 0 rapidly.  [From the continuum form of the Hamiltonian in
(\ref{Hcont}) 
we may estimate $\psi_0(x) \propto e^{-|x|^{3/2}/(3\cdot 2^{3/4})}$.]  
Again we expect the actual boundstate wave-function to have this property 
so $\psi (x,y)$ should be a sufficient approximation in the large $|x|$ 
region.  Therefore we expect $\psi (x,y)$ [with $\epsilon$ given by
(\ref{epsilon})] 
to be a good first approximation to the actual boundstate wave-function at 
small $\lambda$ and the energy of (\ref{Emin})
to be asymptotically correct.  The 
situation is similar to the case of one-dimensional quantum mechanics with 
a weak potential, the single dimension corresponding to $y$. Despite the 
fact that $\psi$ depends on 2 co-ordinates $x$ and $y$, the present problem 
is rather different than the two-dimensional case with a short-range 
potential due to the confining $x$-dependent potential. Essentially, $x$ 
acts as an internal degree of freedom of the magnon whereas $y$ represents
its location.  In particular, the quadratic dependence of the binding 
energy on 
the strength of the potential also occurs in the one-dimensional case.

\begin{figure}
\epsfxsize=\figwidth
\centerline{\epsffile{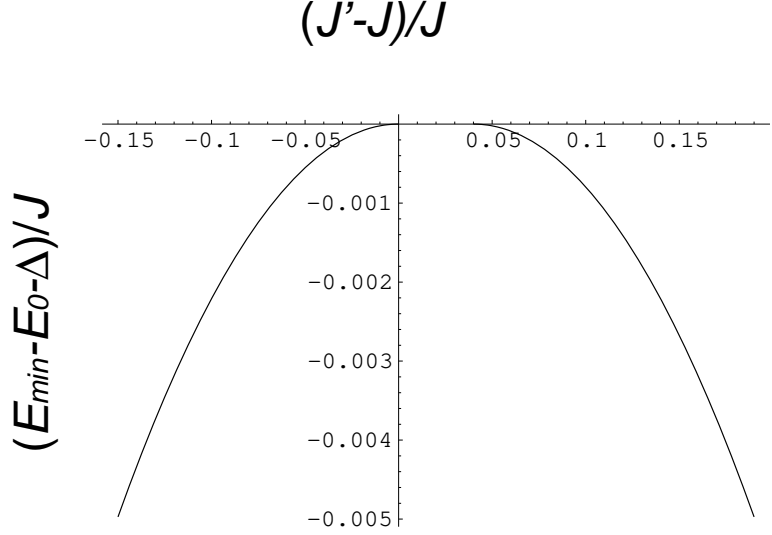}}  
\caption{$E_{min}$ versus $(J'-J)/J$.  The critical value of $J' > J$
for which the energy becomes negative has been assigned an arbitrary value.
The form of the energy dependence for $J'$ larger than this critical
value has been assumed to be the same as the dependence for $J' < J$.}
\label{fig:evl}\end{figure}

Thus, if we assume that $E_{min}$ is not merely
an upper bound on the energy but
actually a good approximation to it, we can plot the energy as a function
of $\lambda = (J'-J)/J$.  This is done in Figure \ref{fig:evl}.  Note that
the critical value of $\lambda > 0$ at which the energy becomes negative
(implying the existence of a bound state) has not been calculated and
so no significance should be attributed to the value indicated on the plot.
Also, the exact form of
the energy dependence for $\lambda>0$ is unknown.  In Figure \ref{fig:evl}
it is plotted with the same quadratic dependence as in the case $\lambda < 0$.
Based on our experience in the previous section with extrapolating 
large-n results to n=2, we might expect the pre-factor of .221 in 
Eq. (\ref{benum}) to only be accurate to within a factor of 2 or so.
However, the fact
that the functional dependence is quadratic in $(J'-J)$ is likely to
be exact, since it is a general result for one-dimensional quantum
systems, as emphasized above.  The numerical results in \cite{Mallwitz}
are probably consistent with quadratic behaviour at small negative
$(J'-J)$ but with a smaller pre-factor.

Therefore, we have shown that in the case of a modified link, in the limit
$J' \to J$,
a bound state exists if and only if $J'<J$.  This results from the fact
that between the antisoliton and soliton there is one more pair of
neighboring sites with no bond between them than  with a double bond.

\acknowledgments
I.A. would like to thank Erik S\o rensen for useful discussions.
This research was supported by NSERC of Canada.

\appendix \section{Calculation of Overlaps Arising in Determination
of True Ground State}
\label{app:overlaps}
\subsection{Calculation of $<0|0>$}
For the symmetric ground state, we first consider just two of the sites.
\begin{eqnarray}
<^{i_1i_2},_{i_2i_3}|_{j_1j_2},^{j_2j_3}>&=&
(\delta^{i_1}_{j_1}\delta^{i_2}_{j_2}+\delta^{i_1}_{j_2}\delta^{i_2}_{j_1})
(\delta^{j_2}_{i_2}\delta^{j_3}_{i_3}+\delta^{j_2}_{i_3}\delta^{j_3}_{i_2})
\nonumber \\
&=&(n+2)\delta^{i_1}_{j_1}\delta^{j_3}_{i_3}
+\delta^{i_1}_{i_3}\delta^{j_3}_{j_1}
\end{eqnarray}
We then consider three sites.
\begin{eqnarray}
<^{i_1i_2},_{i_2i_3},^{i_3i_4}|_{j_1j_2},^{j_2j_3},_{j_3j_4}>&=&
(\delta^{i_1}_{j_1}\delta^{i_2}_{j_2}+\delta^{i_1}_{j_2}\delta^{i_2}_{j_1})
(\delta^{j_2}_{i_2}\delta^{j_3}_{i_3}+\delta^{j_2}_{i_3}\delta^{j_3}_{i_2})
(\delta^{i_3}_{j_3}\delta^{i_4}_{j_4}+\delta^{i_3}_{j_4}\delta^{i_4}_{j_3})
\nonumber \\
&=&[(n+2)\delta^{i_1}_{j_1}\delta^{j_3}_{i_3}
+\delta^{i_1}_{i_3}\delta^{j_3}_{j_1}]
(\delta^{i_3}_{j_3}\delta^{i_4}_{j_4}+\delta^{i_3}_{j_4}\delta^{i_4}_{j_3})
\nonumber \\
&=&[(n+1)(n+2)+1]\delta^{i_1}_{j_1}\delta^{i_4}_{j_4}
+\delta^{i_1}_{j_4}\delta^{i_4}_{j_1}
\end{eqnarray}
We see that this pattern will continue.  That is, for $L$ sites,
with $L$ even, we obtain
\begin{equation}
<0|0>=a_L\delta^{i_1}_{j_1}\delta^{j_{L+1}}_{i_{L+1}}
+\delta^{i_1}_{i_{L+1}}\delta^{j_{L+1}}_{j_1}
\end{equation}
Where $a_L$ is given by the recursion relation
\begin{equation}
a_{L+1}=(n+1)a_L+1
\end{equation}
However, this is simply a geometric series.
\begin{eqnarray}
a_1&=&1 \nonumber \\
a_2&=&(n+1)+1 \nonumber \\
a_3&=&(n+1)^2+(n+1)+1 \nonumber \\
a_L&=&\sum_{i=0}^{L-1}(n+1)^i=[(n+1)^L-1]/n
\end{eqnarray}
Thus for a finite chain of L sites,
\begin{equation}
<0|0>=[((n+1)^L-1)/n]\delta^{i_1}_{j_1}\delta^{j_{L+1}}_{i_{L+1}}
+\delta^{i_1}_{i_{L+1}}\delta^{j_{L+1}}_{j_1}
\end{equation}
If we now wish to impose periodic boundary conditions, this is
equivalent to replacing $i_{L+1}$ by $i_1$ and $j_{L+1}$ by $j_1$
in the above expression.  Doing this, and contracting the indices
yields
\begin{equation}
<0|0>=[((n+1)^L-1)/n]n+n^2=(n+1)^L+n^2-1
\label{0norm}
\end{equation}
Taking $L\to \infty$ first, then large $n$, this may be approximated as:
\begin{equation} <0|0>=(n+1)^L.\end{equation}
\subsection{Calculation of $<EO|EO>$}
To calculate this matrix element, we can use the previous result.  Each
double bond can be thought of as a two site $|0>$ state with periodic
boundary conditions.  Thus, for L even
\begin{equation}
<EO|EO>=[(n+1)^2+n^2-1]^{L/2}=[2n(n+1)]^{L/2}
\label{EOnorm}
\end{equation}

\subsection{Calculation of $<\alpha_r|\alpha_r>$}
We again use our previous results to calculate this matrix element.
The four sites which comprise the soliton-antisoliton pair can be thought
of as a four site $|0>$ state with periodic boundary conditions while the
remainder of the $|\alpha_r>$ state is identical to the $|EO>$ state.
Thus, for $L$ even
\begin{equation}
<\alpha_r|\alpha_r>=[(n+1)^4+n^2-1][2n(n+1)]^{(L-4)/2}
\end{equation}

\subsection{Calculation of $<\beta_r|\beta_r>$}
The overlap of $|\beta_r>$ with itself consists of the overlap of a
two site $|EO>$ state (or equivalently a two site $|0>$ state) with
itself and an $(L-2)$ site $|0>$ state with itself.  Thus, for $L$
even
\begin{equation}
<\beta_r|\beta_r>=2n(n+1)[(n+1)^{L-2}+n^2-1]
\end{equation}

\subsection{Calculation of $<0|\beta_r>$}
We need only consider the four sites which comprise the soliton-antisoliton
pair.  The remainder of the matrix element is identical to $<0|0>$.
\begin{eqnarray}
<^{i_1i_2},_{i_2i_3},^{i_3i_4},_{i_4i_5}|_{j_1j_2},^{j_3j_4},_{j_3j_4},
^{j_2j_5}>&=&
(\delta^{i_1}_{j_1}\delta^{i_2}_{j_2}+\delta^{i_1}_{j_2}\delta^{i_2}_{j_1})
(\delta^{j_3}_{i_2}\delta^{j_4}_{i_3}+\delta^{j_3}_{i_3}\delta^{j_4}_{i_2})
(\delta^{i_3}_{j_3}\delta^{i_4}_{j_4}+\delta^{i_3}_{j_4}\delta^{i_4}_{j_3})
(\delta^{j_2}_{i_4}\delta^{j_5}_{i_5}+\delta^{j_2}_{i_5}\delta^{j_5}_{i_4})
\nonumber \\
&=&(\delta^{i_1}_{j_1}\delta^{j_3}_{j_2}\delta^{j_4}_{i_3}
+\delta^{i_1}_{j_1}\delta^{j_3}_{i_3}\delta^{j_4}_{j_2}
+\delta^{i_1}_{j_2}\delta^{j_3}_{j_1}\delta^{j_4}_{i_3}
+\delta^{i_1}_{j_2}\delta^{j_3}_{i_3}\delta^{j_4}_{j_1}) \nonumber \\
&&(\delta^{i_3}_{j_3}\delta^{j_2}_{j_4}\delta^{j_5}_{i_5}
+\delta^{i_3}_{j_3}\delta^{j_2}_{i_5}\delta^{j_5}_{j_4}
+\delta^{i_3}_{j_4}\delta^{j_2}_{j_3}\delta^{j_5}_{i_5}
+\delta^{i_3}_{j_4}\delta^{j_2}_{i_5}\delta^{j_5}_{j_3})
\nonumber \\
&=&(2n^2+6n+4)\delta^{i_1}_{j_1}\delta^{j_5}_{i_5}
+2(n+1)\delta^{i_1}_{i_5}\delta^{j_5}_{j_1} \nonumber \\
&=&2(n+1)[(n+2)\delta^{i_1}_{j_1}\delta^{j_5}_{i_5}
+\delta^{i_1}_{i_5}\delta^{j_5}_{j_1}]
\end{eqnarray}
We now note that this is precisely $2(n+1)$ times the expression we obtain
from a two site $<0|0>$ aside from a renaming of indices.  Thus, since
the remaining $L-4$ sites of the $|\beta_r>$ state are identical
to those for the $|0>$ state
\begin{equation}
<0|\beta_r>=2(n+1)[(n+1)^{L-2}+n^2-1]\approx [2/(n+1)]<0|0>
\label{0betaoverlap}\end{equation}

\subsection{Calculation of $<EO|\alpha_r>$}
As in the previous calculation, we need only consider the four sites
which comprise the soliton-antisoliton pair.  The remainder of the matrix
element is identical to $<EO|EO>$.
\begin{eqnarray}
<^{i_1i_2},_{i_1i_2},^{i_3i_4},_{i_3i_4}|
_{j_1j_2},^{j_2j_3},_{j_3j_4},^{j_4j_1}>
&=&(\delta^{i_1}_{j_1}\delta^{i_2}_{j_2}+\delta^{i_1}_{j_2}\delta^{i_2}_{j_1})
(\delta^{j_2}_{i_1}\delta^{j_3}_{i_2}+\delta^{j_2}_{i_2}\delta^{j_3}_{i_1})
(\delta^{i_3}_{j_3}\delta^{i_4}_{j_4}+\delta^{i_3}_{j_4}\delta^{i_4}_{j_3})
(\delta^{j_4}_{i_3}\delta^{j_1}_{i_4}+\delta^{j_4}_{i_4}\delta^{j_1}_{i_3})
\nonumber \\
&=&[(2n+2)\delta^{j_3}_{j_1}][(2n+2)\delta^{j_1}_{j_3}] \nonumber \\
&=&4n(n+1)^2
\end{eqnarray}
Recall that for a four site $|EO>$ chain,
\begin{equation}
<EO|EO>=4n^2(n+1)^2
\end{equation}
Thus, adding in the other $L-4$ sites, we obtain
\begin{equation}
<EO|\alpha_r>=(1/n)<EO|EO>=(1/n)[2n(n+1)]^{L/2}\approx (1/n)<EO|EO>
\label{EOalphaoverlap}\end{equation}

\subsection{Calculation of $<\alpha_r|\alpha_s>$ and $<EO|\alpha_r\alpha_s>$
for $|r-s|>3$}
\label{alpharalphas}
In this case, the two soliton-antisoliton pairs do not overlap.  Thus,
comparison with the calculation of $<EO|\alpha_r>$ yields
\begin{equation}
<\alpha_r|\alpha_s>=(1/n^2)[2n(n+1)]^{L/2}\approx (1/n)^2<EO|EO>
\label{alphaoverlap}
\end{equation}

\subsection{Calculation of $<\alpha_t|\alpha_r\alpha_s>$ for
$|r-s|>3,|r-t|>3,|s-t|>3$}
\begin{figure}
\epsfxsize=\figwidth
\centerline{\epsffile{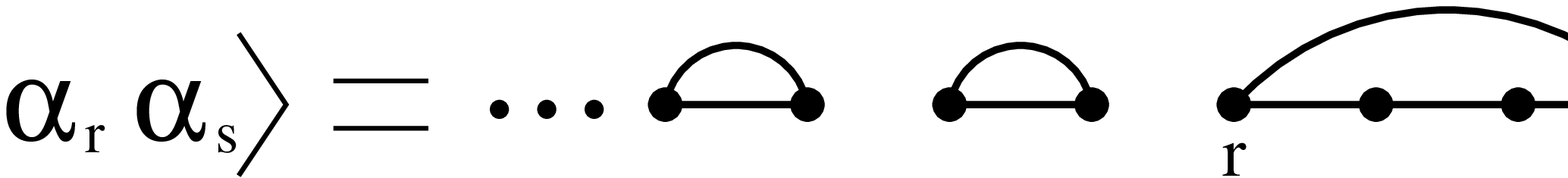}}  
\caption{Double alpha state}
\label{fig:alpha2def}\end{figure}

Here $|\alpha_r\alpha_s>$ is defined in the obvious way in Figure
\ref{fig:alpha2def}.
In this case, none of the soliton-antisoliton pairs overlap.  Thus, comparison
with the calculation of $<EO|\alpha_r>$ yields
\begin{equation}
<\alpha_t|\alpha_r\alpha_s>=(1/n^3)[2n(n+1)]^{L/2}
\end{equation}

\subsection{Calculation of $<\beta_r|\beta_s>$ and $<0|\beta_r\beta_s>$
for $|r-s|>3$}
As in section \ref{alpharalphas}, comparison with the calculation of
$<0|\beta_r>$ yields
\begin{equation}
<\beta_r|\beta_s>=4(n+1)^2[(n+1)^{L-4}+n^2-1]\approx [2/(n+1)]^2<0|0>
\label{betaoverlap}
\end{equation}

\subsection{Calculation of $<\beta_t|\beta_r\beta_s>$ for
$|r-s|>3,|r-t|>3,|s-t|>3$}
\begin{figure}
\epsfxsize=\figwidth
\centerline{\epsffile{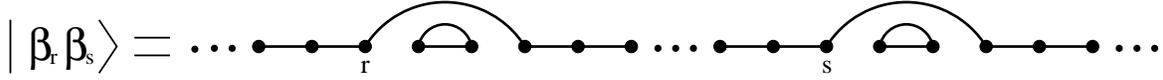}}  
\caption{Double beta state}
\label{fig:beta2def}\end{figure}

Here $|\beta_r\beta_s>$ is defined in the obvious way in Figure
\ref{fig:beta2def}.
Comparison with the calculation of $<0|\beta_r>$ yields
\begin{equation}
<\beta_t|\beta_r\beta_s>=8(n+1)^3[(n+1)^{L-6}+n^2-1]
\end{equation}

\section{Variational Verification of True Ground State}
\label{app:variation}
We consider states of the form
\begin{eqnarray}
|\psi_1>&=&|0>+a|\beta> \nonumber \\
|\psi_2>&=&|EO>+b|\alpha>
\end{eqnarray}
Where
\begin{eqnarray}
|\beta> &\equiv& (1/L)\sum_r|\beta_r> \nonumber \\
|\alpha> &\equiv& [1/(2L)]\sum_{r}|\alpha_{2r}>
\end{eqnarray}
We will see that variations in the parameters $a$ and $b$ do not affect
the energies of the states at order $1/n$ thus verifying that the symmetric
ground state is the true ground state (in the large $n$ limit).  Throughout
this section, we will make the assumption that for large $L$, when the
Hamiltonian acts on $|\alpha_r>$ or $|\beta_r>$, we can neglect the terms
of the Hamiltonian acting on the five sites $r-1$ to $r+3$ since their effect
is suppressed by a factor of $L$ compared to the other sites (of which
there are $L-5$).

For the symmetric ground state, we wish to calculate
\begin{equation}
E_1=<\psi_1|H|\psi_1>/<\psi_1|\psi_1>
\end{equation}
We already know $H|0>$.  Now, for $|r-s|>3$,
\begin{equation}
H_s|\beta_r>=-(1+2/n)|\beta_r>-(1/n)|\beta_r\beta_s>
\end{equation}
Where $|\beta_r\beta_s>$ is defined in Appendix \ref{app:overlaps}.
Thus, using our large $L$ assumption,
\begin{equation}
H|\beta_r>=-L(1+2/n)|\beta_r>-(1/n)\sum_s|\beta_r\beta_s>
\end{equation}
Thus, we see that the normalization conditions ((\ref{0norm}), 
(\ref{0betaoverlap}) and
(\ref{betaoverlap})) and the action of $H$ are consistent with
\begin{equation}
|\beta> \approx [2/(n+1)]|0>
\end{equation}
Although this equation is not literally true, the variational calculations
give the same result as if it were true.  Therefore, $|\psi_1>$ just acts
as a scalar multiple of $|0>$ and so
\begin{equation}
<\psi_1|H|\psi_1>/<\psi_1|\psi_1>=<0|H|0>/<0|0>=-L(1+2/n)
\end{equation}
to first order in $1/n$.

For the non-symmetric ground state, we need to calculate
\begin{equation}
E_2=<\psi_2|H|\psi_2>/<\psi_2|\psi_2>
\end{equation}
Calculations analogous to those for the symmetric yield groundstate demonstrate
that the normalization conditions ((\ref{EOnorm}), (\ref{EOalphaoverlap})
 and (\ref{alphaoverlap}))
and the action of $H$ are consistent with
\begin{equation}
|\alpha> \approx (1/n)|EO>
\end{equation}
Again, this equation is not literally true; the variational calculations just
give the same result as if it were true.  Therefore, $|\psi_2>$ just acts
as a scalar multiple of $|EO>$ and so
\begin{equation}
<\psi_2|H|\psi_2>/<\psi_2|\psi_2>=<EO|H|EO>/<EO|EO>=-L(1+1/n)
\end{equation}
to first order in $1/n$.

Thus, variations in $a$ and $b$ do not affect the energies of the states
$|\psi_1>$ and $|\psi_2>$ at order $1/n$ as claimed.

\section{Calculation of Overlaps for Soliton-Antisoliton States}
\label{app:solitonoverlaps}
In this section, we will only consider states in which the soliton is to the
left of the antisoliton (that is, $|\widetilde{r,s}>$ for which $s>r$).
The other cases
are easily seen to produce analogous results.
In the forthcoming calculations, we will find the overlap of 2-site
wave functions, one corresponding to a single bond and the other to a double
bond, to be useful.  This is just
\begin{equation}
<^{i_1i_2},_{i_1i_2}|_{j_1j_2},^{j_2j_3}>=
(\delta^{i_1}_{j_1}\delta^{i_2}_{j_2}+\delta^{i_1}_{j_2}\delta^{i_2}_{j_1})
(\delta^{j_2}_{i_1}\delta^{j_3}_{i_2}+\delta^{j_2}_{i_2}\delta^{j_3}_{i_1})
=2(n+1)\delta^{j_3}_{j_1}
\end{equation}
This result can obviously be extended to longer chains.
\begin{equation}
<^{i_1i_2},_{i_1i_2}...^{i_{x-2}i_{x-1}},_{i_{x-2}i_{x-1}}|_{j_1j_2},^{j_2j_3}...
_{j_{x-2}j_{x-1}},^{j_{x-1}j_x}>=[2(n+1)]^{(x-1)/2}\delta^{j_x}_{j_1}
\label{EOOoverlap}
\end{equation}
Where we note that in the above, $x$ must be odd.

\begin{figure}
\epsfxsize=\figwidth
\centerline{\epsffile{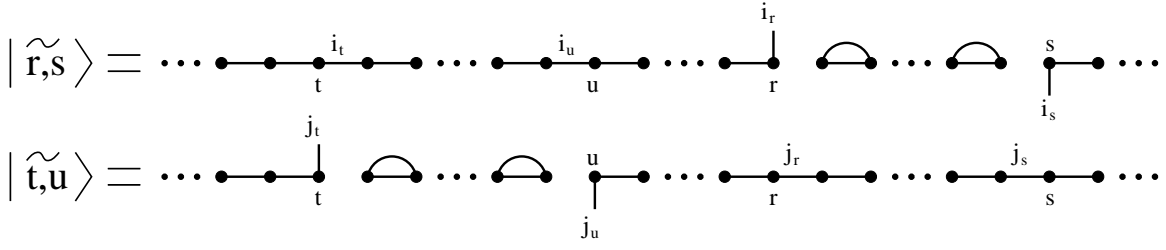}}  
\caption{Site and index labeling when $t<u\le r<s$.}
\label{fig:tursoverlap}
\end{figure}

Let us first calculate the overlap $<\widetilde{r,s}|\widetilde{t,u}>$
when $t<u\le r<s$.
We label indices as in Figure \ref{fig:tursoverlap}.
We define $x'=u-t$, $x=s-r$ and $d=r-u+1$.
From (\ref{EOOoverlap}), the overlap
for sites $t+1$ through $u-1$ is
\begin{equation}
[2(n+1)]^{(x'-1)/2}\delta^{i_u}_{i_t}
\end{equation}
Similarly, the overlap for sites $r+1$ through $s-1$ is
\begin{equation}
[2(n+1)]^{(x-1)/2}\delta^{j_r}_{j_s}
\end{equation}
From Appendix \ref{app:overlaps} we know that the overlap for sites $u$ through $r$,
\begin{equation}
[((n+1)^d-1)/n]\delta^{j_u}_{i_u}\delta^{i_r}_{j_r}+\delta^{j_u}_{j_r}
\delta^{i_r}_{i_u}
\end{equation}
and the overlap for sites $s$ through $t$ (periodic boundary conditions) is
\begin{equation}
[((n+1)^{L-d-x-x'+2}-1)/n]\delta^{j_s}_{i_s}\delta^{i_t}_{j_t}
+\delta^{j_s}_{j_t}\delta^{i_t}_{i_s}
\end{equation}
Therefore, the entire overlap is
\begin{equation}
[2(n+1)]^{(x+x'-2)/2}\delta^{i_u}_{i_t}\delta^{j_r}_{j_s}[(((n+1)^d-1)/n)
\delta^{j_u}_{i_u}\delta^{i_r}_{j_r}+\delta^{j_u}_{j_r}\delta^{i_r}_{i_u}]
[(((n+1)^{L-d-x-x'+2}-1)/n)\delta^{j_s}_{i_s}\delta^{i_t}_{j_t}
+\delta^{j_s}_{j_t}\delta^{i_t}_{i_s}]
\end{equation}
Now, for our purposes, $i_r=j_t$ and $i_s=j_u$ since the two states
considered here are intended to be the same state either translated or
acted on by our Hamiltonian (or both) and both of these operations
preserve the value of the free index.  Now, in order to avoid mixing with
the singlet states, we also assume $i_r\ne i_s$ and $j_t\ne j_u$.  Thus,
after expanding and contracting indices, the above expression reduces to
\begin{equation}
[2(n+1)]^{(x+x'-2)/2}[((n+1)^d-1)/n+((n+1)^{L-d-x-x'+2}-1)/n]
\end{equation}
In the large $L$ and large $n$ limits (taking $L$ large first
as usual), this reduces to
\begin{equation}
2^{(x+x'-2)/2}n^{L-d-(x+x')/2}
\end{equation}

\begin{figure}
\epsfxsize=\figwidth
\centerline{\epsffile{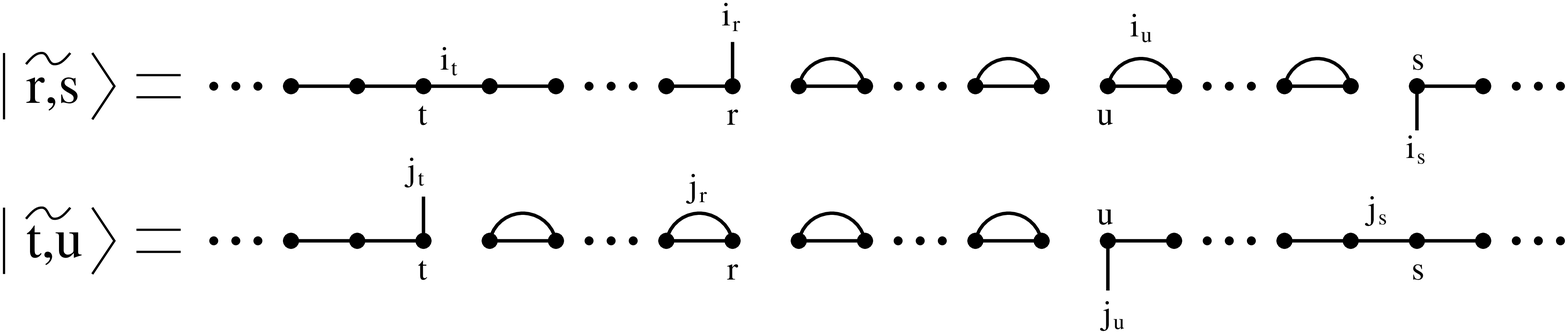}}  
\caption{Site and index labeling when $t<r\le u<s$.}
\label{fig:trusoverlap}
\end{figure}

We now consider the case in which $t<r\le u<s$.  We label the sites and
indices as in Figure \ref{fig:trusoverlap}.
We define $x'=u-t$, $x=s-r$ and $d=u-r-1$.  Routine calculations similar to
those above demonstrate that the overlap is equal to
\begin{equation}
[2(n+1)]^{(x+x'-2)/2-d}\delta^{i_r}_{i_t}\delta^{j_u}_{j_s}[2n(n+1)]^{d/2}
[(((n+1)^{L-x-x'+d+2}-1)/n)\delta^{j_s}_{i_s}\delta^{i_t}_{j_t}
+\delta^{j_s}_{j_t}\delta^{i_t}_{i_s}]
\end{equation}
Using the same conditions on the indices as above, the overlap reduces to
\begin{equation}
[2(n+1)]^{(x+x'-2)/2-d}[2n(n+1)]^{d/2}[((n+1)^{L-x-x'+d+2}-1)/n]
\end{equation}
In the large $L$ and large $n$ limits (taking $L$ large first as usual),
the overlap reduces to
\begin{equation}
2^{(x+x'-d-2)/2}n^{L+d-(x+x')/2}
\end{equation}

\begin{figure}
\epsfxsize=\figwidth
\centerline{\epsffile{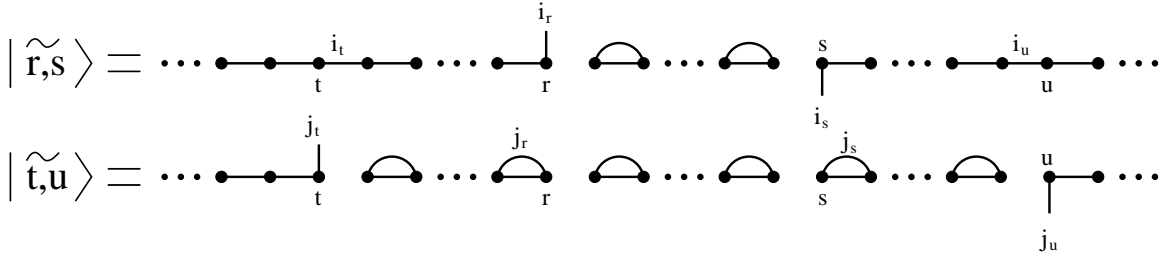}}  
\caption{Site and index labeling when $t<r\le s<u$.}
\label{fig:trsuoverlap}
\end{figure}

Lastly, we must consider the case where $t<r\le s<u$.  We label the indices
as in Figure \ref{fig:trsuoverlap}.
We define $x'=u-t$, $x=s-r$ and $d=u-s$.  Routine calculations similar to
those show above demonstrate that the overlap is equal to
\begin{equation}
[2(n+1)]^{(x'-x)/2}\delta^{i_r}_{i_t}\delta^{i_u}_{i_s}[2n(n+1)]^{(x-1)/2}
[(((n+1)^{L-x'}-1)/n)\delta^{j_u}_{i_u}\delta^{i_t}_{j_t}
+\delta^{j_u}_{j_t}\delta^{i_t}_{i_u}]
\end{equation}
Using the same conditions on the indices as above, this reduces to
\begin{equation}
[2(n+1)]^{(x'-x)/2}[2n(n+1)]^{(x-1)/2}[(((n+1)^{L-x'}-1)/n)]
\end{equation}
In the large $L$ and large $n$ limits (taking $L$ large first as usual),
the overlap reduces to
\begin{equation}
2^{(x'-1)/2}n^{L+(x-x')/2-1}
\end{equation}

Thus, we see that the overlap $<\widetilde{r,s}|\widetilde{t,u}>$ is order
$n^{L-1}$ if $r=t$, $s=u$ and higher order in $1/n$ in every other case.
Although, we have only dealt with the cases in which $s>r$ and $u>t$, it
can easily be seen that the other cases yield the same conclusion.

\section{Positivity of $\psi_0(\lowercase{x})$}
\label{app:positivity}
We will now prove that $\psi_0(x)$ is positive for all odd $x$.  Since we know
$\psi_0(x)$ is an even function of $x$, we need only prove the result for
$x\ge 1$.  In order for $\psi_0$ to be normalizable, $\psi_0(x)$ must converge
to zero.
Now, by construction, $\psi_0(1)>0$.  Assume $\psi_0(x)>0 \ \forall x$ such
that $1\le x\le X$, with
$x,X$ odd. We will demonstrate that $\psi(X+2)>0$ by contradiction.  Thus, by
induction, $\psi_0(x)>0 \ \forall x$ odd.

Assume $\psi(X+2)<0$.  Then, by (\ref{recursion2})
\begin{eqnarray}
\psi_0(X+4)&=&[(X+2-n\delta E-1)/(2\sqrt2)]\psi_0(X+2)-\psi_0(X) \nonumber \\
\Rightarrow \psi_0(X+4)-\psi_0(X+2)&=&
[(X+1-2\sqrt2-n\delta E)/(2\sqrt2)]\psi_0(X+2)-\psi_0(X) < 0 \nonumber \\
\Rightarrow \psi(X+4) &<& \psi(X+2) < 0
\end{eqnarray}
Since $X+1-2\sqrt2-n\delta E>2-2\sqrt2-n\delta E>0$,
$\psi_0(X+2)<0$ and $\psi_0(X)>0$.  Now,
if $\psi_0(x+2)<\psi_0(x)<0$ for $x\ge 3$, then
\begin{eqnarray}
\psi_0(x+4)&=&[(x+2-n\delta E-1)/(2\sqrt2)]\psi_0(x+2)-\psi_0(x) \nonumber \\
\Rightarrow \psi_0(x+4)-\psi_0(x+2)&=&
[(x+1-2\sqrt2-n\delta E)/(2\sqrt2)]\psi_0(x+2)-\psi_0(x) \nonumber \\
\Rightarrow \psi_0(x+4)-\psi_0(x+2)&\le&
\psi_0(x+2)-\psi_0(x) <0 \nonumber \\
\Rightarrow \psi(x+4) &<& \psi(x+2) < 0 \nonumber \\
\end{eqnarray}
Since $x+1-2\sqrt{2}-n\delta E>4-2\sqrt{2}-n\delta E > 2\sqrt2$.
Thus, by induction, $\psi_0(x)\le \psi_0(X+2)<0 \ \forall x\ge X+2$
which contradicts the fact that $\psi_0(x) \to 0$ as $x \to \infty$.
Therefore, we must have that $\psi_0(X+2)>0$ which completes our initial
induction and so $\psi_0(x)>0$ for all $x$.
Here we have used the numerically determined value of $n\delta E\approx
-3.747$.  Higher energy eigenfunctions will not be positive.

\end{document}